\documentclass[sigplan,nonacm]{acmart}

\usepackage{listings}

\usepackage[table]{xcolor}
\usepackage{caption}

\definecolor{dbgreen}{RGB}{44,160,44}
\definecolor{promptblue}{RGB}{31,119,180}
\definecolor{llmred}{RGB}{214,39,40}
\definecolor{evalorange}{RGB}{255,127,14}
\definecolor{codebg}{RGB}{245,245,245}

\usepackage{tikz}
\usepackage{amsmath}
\usepackage{amsfonts}
\usepackage{amsmath}

\usepackage{filecontents}
\usepackage{amsfonts}

\usepackage{cleveref}
\crefformat{section}{\S#2#1#3}
\Crefformat{section}{\S#2#1#3}
\usepackage[english]{babel}
\usepackage{enumitem }
\usepackage{blindtext}
\usepackage{color}
\usepackage{xspace}
\usepackage{algorithm}
\usepackage[noend]{algpseudocode}
\usepackage{multirow}
\usepackage{subcaption}
\usepackage{listings}
\usepackage{tabularx}
\usepackage{booktabs}
\usepackage{quantikz}

\usepackage{graphicx}
\usepackage{gensymb}
\usepackage{flushend}
\usepackage{url}

\newcommand{\GPTF}{\textsc{GPT-5-mini}\xspace}
\newcommand{\GPTFC}{\textsc{GPT-5.3-Codex}\xspace}
\newcommand{\ClaudeOpus}{\textsc{Claude Opus 4.6}\xspace}
\newcommand{\ClaudeSonnet}{\textsc{Claude Sonnet 4.6}\xspace}
\newcommand{\GeminiTP}{\textsc{Gemini 3 Pro}\xspace}
\newcommand{\GeminiTF}{\textsc{Gemini 3 Flash}\xspace}

\usepackage[skip=3pt]{caption}
\usepackage[most]{tcolorbox}

\newcommand{\name}{$\sf{QSA}$\xspace}	

\usepackage{amssymb}

\graphicspath{{Figures/}}

\begin{document}

\newcommand{\BULLET}{\vspace{+.00in} \noindent $\bullet$ \hspace{+.00in}}

\newcommand{\etc}{\emph{etc.}\xspace}
\newcommand{\ie}{\emph{i.e.,}\xspace}
\newcommand{\eg}{\emph{e.g.,}\xspace}
\newcommand{\etal}{\emph{et al.}\xspace}
\newcommand{\wrt}{\emph{w.r.t.}\xspace}
\newcommand{\aka}{\emph{a.k.a.}\xspace}

\title{Toward Quantum Software Automation: A Quantum-Aware Harness for LLM-Guided Evolution}


\author{%
Lily Jiaxin Wan, Deming Chen, Klara Nahrstedt, and Bo Chen
}

\affiliation{%
  \institution{University of Illinois Urbana-Champaign}
  \city{Urbana}
  \state{Illinois}
  \country{USA}
}

\begin{abstract}
Quantum software is critical for improving the efficiency and reliability of scarce quantum hardware.
However, its design still relies heavily on ad-hoc, handcrafted heuristics that are often suboptimal and quickly become obsolete as quantum hardware evolves.
LLM-guided evolutionary search offers a promising way to automatically explore complex software designs, but existing search frameworks lack the quantum-specific support needed for efficient evolution: verification is expensive, feedback is sparse, and heterogeneous quantum programs require different optimization objectives.
In this paper, we present \name, a \emph{quantum-aware harness for LLM-guided evolutionary search} 
toward automating quantum software design.
\name equips the search with three forms of quantum-specific guidance: an evolution-hardness-guided coreset and approximate scoring to reduce verification cost, static and snapshot analyses to provide fine-grained execution context, and task-specific rewards for compiler passes and runtime policies.
We evaluate \name on the IBM Quantum platform across three benchmark suites.
For multiprogramming, \name improves QPU utilization by 4.2\%--9.5\% and Hellinger fidelity by 15.2\%--19.5\% over the state of the art.
For error mitigation, \name reduces mitigation
time by at least 96.8\% while achieving comparable or better fidelity.
These gains require only \$6.9 in LLM API cost over 11.3 hours.

\end{abstract}

\maketitle

\section{Introduction}
\label{sec:introduction}

\begin{figure}
    \centering
    \includegraphics[width=.65\linewidth]{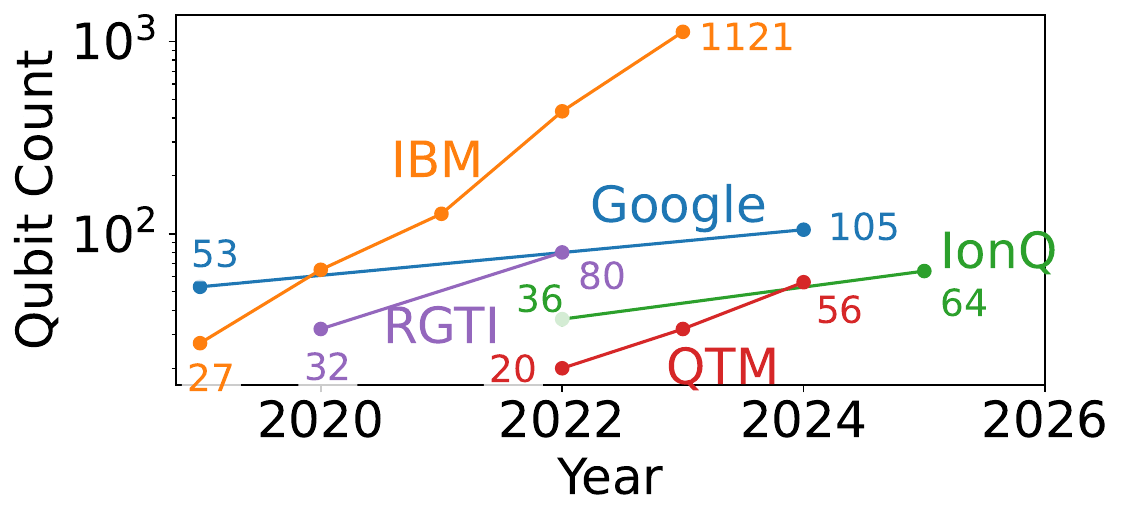}
    \caption{Physical qubit count of major providers. }
    \label{fig:qubit_number_results}
\end{figure}

\begin{figure}[t]
    \centering
    \includegraphics[width=.8\linewidth]{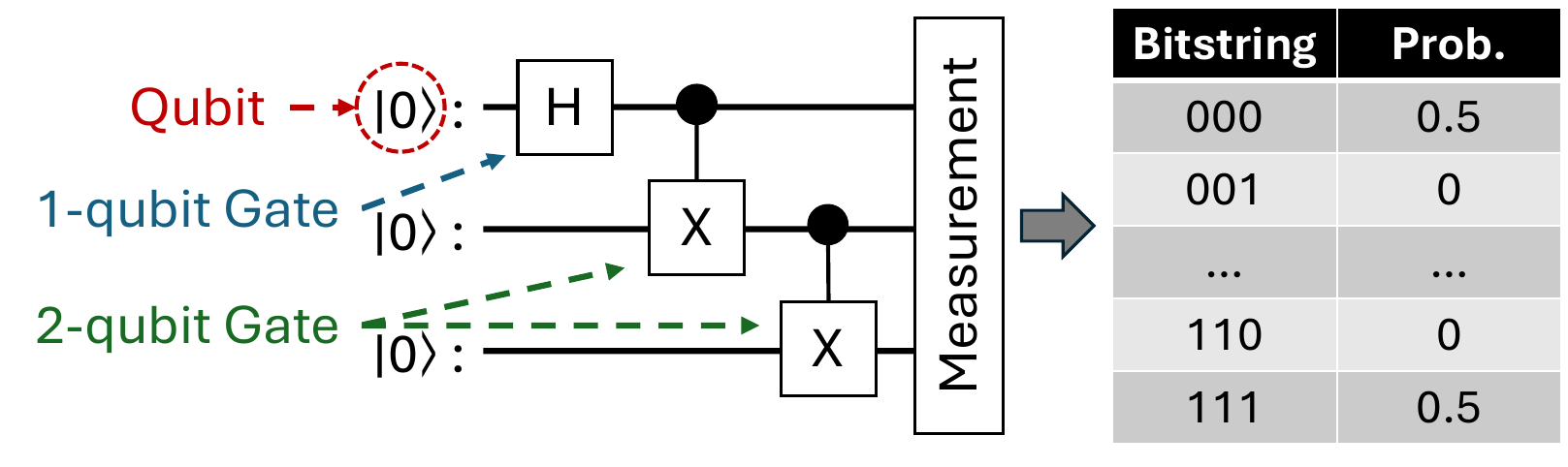}
    \caption{An example GHZ~\cite{greenberger1989going} quantum circuit. The circuit consists of three qubits, one 1-qubit gate, and two 2-qubit gates. The measurement module produces classical outcome of 3-bit bitstrings with different probability.}
    \label{fig:circuit_example}
\end{figure}

Advances in quantum computing have enabled quantum-as-a-service through cloud platforms~\cite{ibm_quantum_platform,aws_braket,quantinuum,ionq,azure_quantum,rigetti_qcs}. 
As shown in Figure~\ref{fig:qubit_number_results}, the scale of quantum hardware, measured by the number of physical quantum bits (qubits), has grown rapidly in recent years.
For instance, IBM increased its physical qubit count by two orders of magnitude from 2021 to 2023.
These advances bring quantum applications closer to reality~\cite{kim2023evidence}. 
A quantum system provides such a service by executing quantum circuits and returning classical outcomes (\eg bitstring probabilities) through sampling.
Figure~\ref{fig:circuit_example} illustrates a 3-qubit GHZ circuit~\cite{greenberger1989going} and its output.

{\em Quantum software} is crucial to modern quantum computing.
It spans \textit{compiler passes} that transform quantum circuits before execution and \textit{runtime policies} that determine how circuits are scheduled on QPUs.
In this work, we study two representative components: error mitigation as a compiler pass and multiprogramming as a runtime policy.
Designing such software, however, is costly.
First, the design space is enormous.
For example, multiprogramming must decide how circuits are grouped for concurrent execution. Partitioning (n) quantum programs into pairs, where (n) is even, already yields
\(\frac{n!}{2^{n/2}\left(\frac{n}{2}\right)!}\)
possible combinations; even ten programs admit 945 different groupings.
Second, evaluating a design typically requires costly quantum simulation or hardware execution, making the design space difficult to model and explore efficiently.
Finding a good design therefore requires extensive trial and error.
As a result, existing quantum software often relies on ad-hoc, handcrafted heuristics (\eg brute-force search and empirical hyperparameter choices), which can be substantially suboptimal.
For instance, our study shows that an ad-hoc multiprogramming approach~\cite{giortamis2025qos} achieves 45.7\% lower fidelity than the offline optimal.
%
Moreover, rapid hardware evolution (Figure~\ref{fig:qubit_number_results}) requires heuristic retuning.

A natural direction is to automate this design process through parameter search. 
Traditional optimization techniques, such as Bayesian optimization~\cite{snoek2012practical,frazier2018tutorial}, reinforcement learning~\cite{mirhoseini2017device,zhou2020transferable}, and program synthesis~\cite{gulwani2017program,solar2008program}, 
typically operate over predefined search spaces, limiting structural code exploration.
%
Recent LLMs provide a more flexible search operator because they can directly generate and revise source code~\cite{lin2025byos,park2025principles,cummins2024meta}. This has led to LLM-guided evolutionary search~\cite{cruzbenito2026evolutionary,novikov2025alphaevolve,openevolve}, where an LLM repeatedly mutates a program, verifies the generated candidate on the target system, and uses the observed reward to guide subsequent mutations. By incorporating system feedback into later iterations, the search can progressively refine 
both program structure and decision logic.
%
However, existing frameworks provide only a generic evolutionary loop and lack the quantum-specific mechanisms needed to interact effectively with quantum systems, 
making them inefficient.
%

This paper presents \name, a \emph{quantum-aware harness for LLM-guided evolutionary search} 
for quantum compiler and runtime optimization.
%
\name equips the search agent with quantum-specific mechanisms that determine \emph{what to verify and how to score it}, \emph{what execution context to expose}, and \emph{how to reward heterogeneous quantum programs}.
Specifically, \name addresses the following three challenges.

\vspace{0.02in}
\noindent\textbf{Challenge 1: Expensive quantum-system verification (\cref{sec:verifier}).}
Verification lies on the critical path of evolutionary search: each candidate program must be evaluated across multiple quantum circuits.
Unfortunately, key metrics such as Hellinger fidelity~\cite{bhattacharyya1943measure} require costly QPU execution or quantum-circuit simulation, making verification cost prohibitive.
%
%
Our key idea is to reduce both \emph{what to verify} and \emph{how to score it}.
First, we develop evolution-hardness-guided coreset extraction.
It repeatedly evolves programs using uniformly sampled verification sets,
evaluates the resulting programs across circuit scales, and identifies the
scale with the lowest post-evolution reward.
We then sample circuit types at this challenging scale to construct the coreset.
Second, we replace expensive fidelity evaluation during evolution with lightweight proxy scores derived from circuit properties.
These techniques substantially reduce verification overhead while preserving effective guidance for evolution.

\vspace{0.02in}
\noindent\textbf{Challenge 2: Sparse feedback for LLM-guided search (\cref{sec:prompter}).}
Generic evolutionary search provides the LLM with program code and final rewards, but little quantum-specific context about the workload or how the program affects it during execution.
Such sparse feedback makes it difficult for the LLM to understand why a mutation succeeds or fails and to generate targeted improvements.
Our key idea is therefore to expose 
quantum-specific execution context.
First, static analysis summarizes quantum-circuit characteristics before execution.
Second, snapshot analysis captures intermediate system states as the evolved program executes.
These signals connect circuit characteristics and program decisions to execution outcomes, enabling more informed mutations and more efficient search.

\vspace{0.02in}
\noindent\textbf{Challenge 3: Heterogeneous programs in quantum software (\cref{sec:reward_design}).}
Quantum software programs can be broadly categorized according to how they interact with quantum circuits.
\emph{Compiler passes} directly transform the circuit, whereas 
\emph{runtime policies} leave the circuit unchanged and instead optimize its execution on quantum hardware.
Although quantum metrics such as fidelity can directly evaluate compiler passes, they are inadequate for runtime policies because they capture policy quality only indirectly.
%
Our key idea is therefore to tailor \emph{how to reward} different quantum programs.
For compiler passes, we derive rewards directly from quantum-system metrics.
For runtime policies, we design a policy reward that jointly captures Pareto optimality and 
global ranking qualities.
%
We measure Pareto optimality across conflicting objectives using ranks from non-dominated sorting~\cite{deb2002fast}, and generalizability using the correlation between the policy's predicted ranks and the corresponding ground-truth ranks.

We implement \name on a Linux server with an AMD Ryzen 9 CPU and evaluate it
using the IBM Quantum platform and public LLM APIs.
For multiprogramming, \name improves QPU utilization by 4.2\%--9.5\% while
increasing fidelity by 15.2\%--19.5\% over the state of the art.
For error mitigation, \name reduces execution time by at least 96.8\% while
achieving comparable or better fidelity.
%
The evolved programs further maintain consistent performance across circuit types and scales.
These gains incur only \$6.9 in LLM API cost over 11.3 hours.

Overall, we make the following contributions:
\begin{enumerate}[leftmargin=*]
    \item We present \name, a quantum-aware harness that enables LLM-guided evolutionary search to automatically 
    optimize quantum compiler passes and runtime policies.
    \item We design three quantum-specific harness mechanisms: an evolution-hardness-guided coreset and approximate scoring for efficient verification, a quantum-aware prompter for fine-grained search feedback, and task-specific rewards for heterogeneous quantum programs.
    \item We evaluate \name on the IBM Quantum platform and demonstrate significant improvements in QPU utilization, fidelity, and 
    mitigation
    time over SOTA systems.
\end{enumerate}

\section{Background and Motivation}

\begin{figure}
    \centering
    \includegraphics[width=1\linewidth]{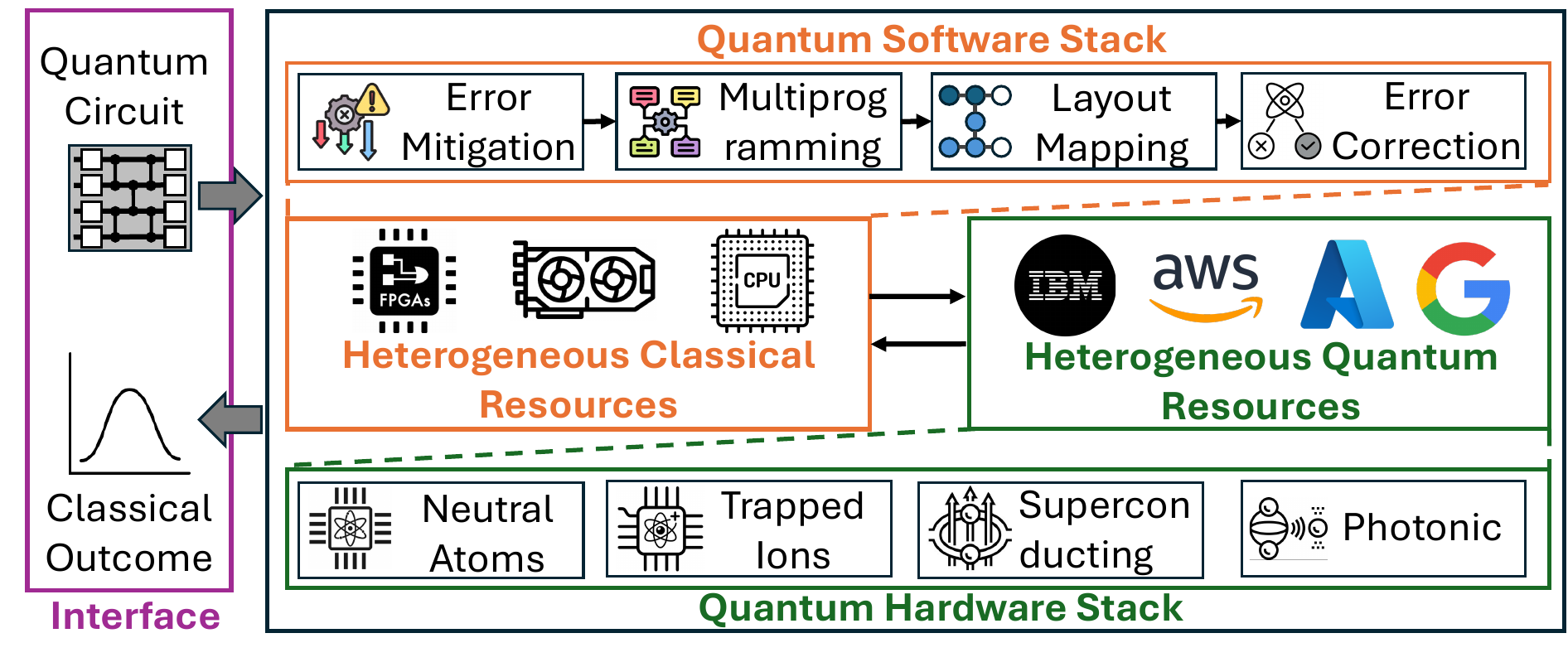}
    \caption{Quantum hybrid computational model.}
    \label{fig:hybrid_model}
    \vspace{-3pt}
\end{figure}
\subsection{Quantum Background}
We introduce the quantum circuit and the quantum hybrid computational model, followed by the key metric.

\vspace{0.02in}
\noindent\textbf{Quantum circuit.}
A quantum circuit, often used to represent a quantum program, is a fundamental computational abstraction executed on quantum hardware.
For example, Figure~\ref{fig:circuit_example} shows a GHZ circuit operating on three qubits, the quantum counterparts of classical bits.
Unlike a classical bit, which takes a value of either 0 or 1, a qubit can be prepared in a superposition of $\ket{0}$ and $\ket{1}$.
In this circuit, a Hadamard ($H$) gate~\cite{shepherd2006role} places the first qubit into superposition.
The subsequent two-qubit controlled-NOT (CNOT or CX) gates~\cite{plantenberg2007demonstration} propagate this superposition across the remaining qubits, creating a three-qubit entangled state.
Measurement maps the quantum state to a classical bitstring, with each possible outcome occurring according to its measurement probability.
Repeated circuit executions, commonly called \emph{shots}, are used to estimate the resulting probability distribution.
Ideally, the GHZ circuit prepares
$(\ket{000}+\ket{111})/\sqrt{2}$; therefore, only \texttt{000} and \texttt{111} have nonzero probabilities, each equal to $0.5$.
More generally, the state of an $n$-qubit system is represented in a $2^n$-dimensional state space, allowing quantum algorithms to exploit superposition, interference, and entanglement for certain computational problems.

\begin{figure}[t]
  \begin{minipage}[t]{0.495\linewidth}
    \centering
    \includegraphics[width=\linewidth]{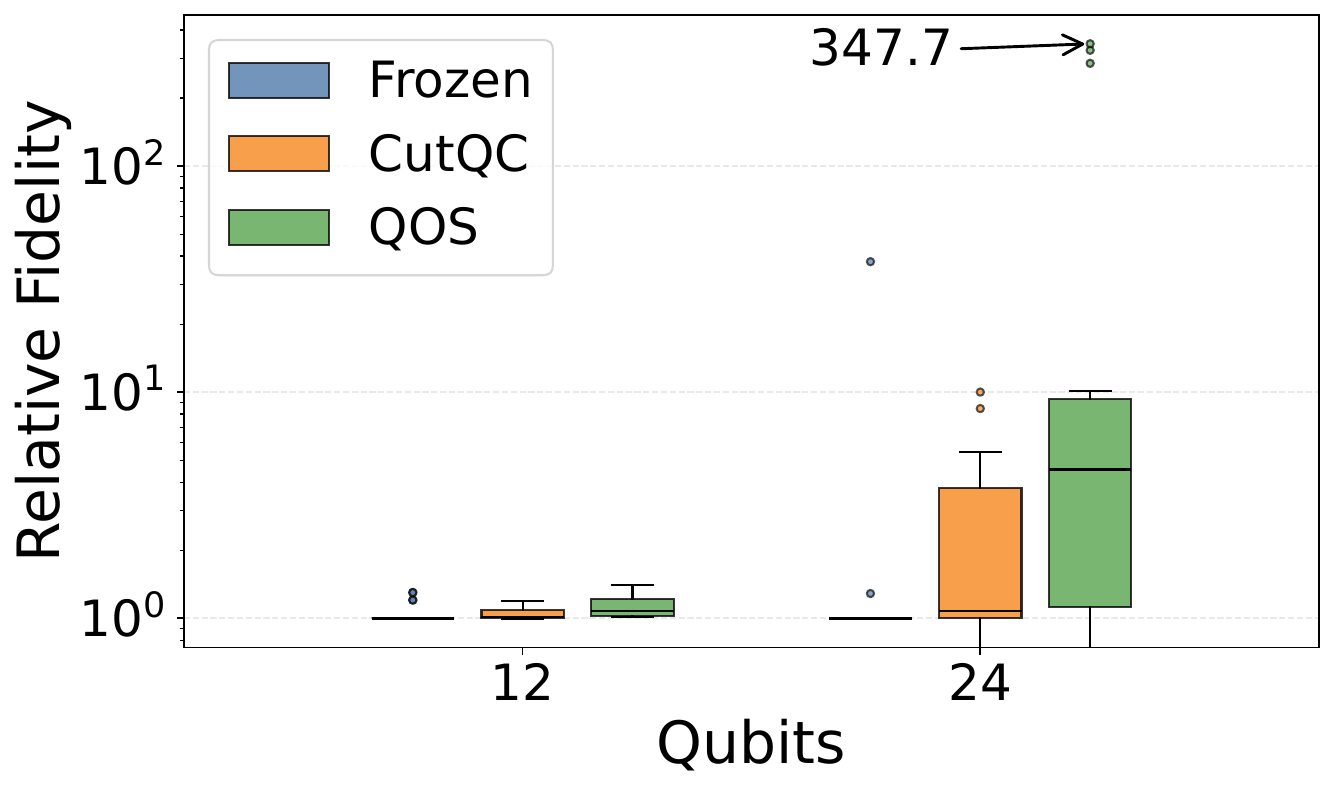}
    \caption{Fidelity results.}
    \label{fig:fidelity_comparison}
  \end{minipage}
  \hfill
  \begin{minipage}[t]{0.495\linewidth}
    \centering
    \includegraphics[width=\linewidth]{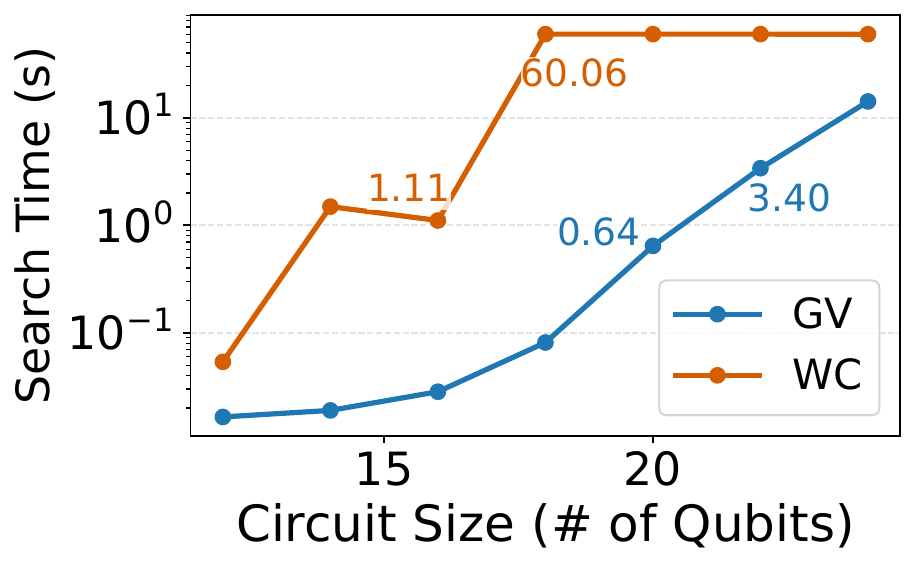}
    \caption{Cut time analysis.}
    \label{fig:cut_pass_timing}
  \end{minipage}
\end{figure}
\vspace{0.02in}
\noindent\textbf{Hybrid computational model.}
Modern quantum computing follows a hybrid execution model that tightly couples {\emph{quantum hardware}} with {\emph{quantum software}}, as illustrated in Figure~\ref{fig:hybrid_model}.
Quantum hardware is realized using four major technologies: neutral atoms~\cite{saffman2016quantum}, trapped ions~\cite{bruzewicz2019trapped}, superconducting circuits~\cite{kjaergaard2020superconducting}, and photonic systems~\cite{slussarenko2019photonic}. 
It executes a circuit by applying a sequence of quantum gates to qubits through physical control operations, evolving the quantum state until measurements produce classical outputs. 
Quantum software compiles programs by mapping circuits onto the target QPU layout~\cite{kirmemics2025weaver}, and mitigates noise either before execution~\cite{ayanzadeh2023frozenqubits,tang2021cutqc,giortamis2025qos} or after execution~\cite{li2017efficient,temme2017error} using classical computing resources such as FPGAs, GPUs, and CPUs. 
In this hybrid computational model, a user submits the design of a quantum circuit, the quantum hardware and software collaborate to output classical outcomes.
%

\vspace{0.02in}
\noindent\textbf{Quantum system metric.}
We evaluate circuit execution quality on NISQ hardware using \emph{Hellinger fidelity}, which measures the similarity between the ideal output distribution and the noisy distribution observed from hardware execution. Let $p = \{p_i\}$ denote the ideal distribution and $q = \{q_i\}$ the measured distribution. The Hellinger fidelity is defined as
\begin{equation}
\label{sec:fidelity_definition}
F_H(p, q) = \left( \sum_i \sqrt{p_i \, q_i} \right)^2.
\end{equation}
$F_H(p,q) \in [0,1]$, where $F_H=1$ indicates perfect agreement. This metric captures the cumulative impact of noise, gate errors, and decoherence on quantum hardware.
We refer to it as fidelity for brevity.

\subsection{Why Quantum Software Design Matters?}
\label{sec:quantum_software_design}

Quantum software is critical for addressing two major limitations of modern
quantum computing: resource scarcity and computation noise.

\vspace{0.02in}
\noindent\textbf{Resource scarcity.}
Due to their high cost and system complexity, QPUs remain extremely scarce
resources.
Today's public quantum clouds (\eg IBM Quantum~\cite{ibm_quantum_platform},
AWS Braket~\cite{aws_braket}, IonQ~\cite{ionq}, and
Quantinuum~\cite{quantinuum}) expose only a limited number of QPUs and
typically execute one circuit per QPU at a time to reduce interference.
This practice constrains the throughput of quantum services.
Our measurements on IBM Quantum further reveal severe queueing: for example,
the 99.9th-percentile queue length of IBM\_fez reaches 39{,}985 jobs.
Detailed QPU availability and queue statistics are provided in
Appendix~\ref{sec:ibm_logging}.
To improve throughput, recent works explore \emph{multiprogramming}
~\cite{das2019case,tao2025quantum,giortamis2025qos,crovella1991multiprogramming},
which bundles compatible circuits from the same workload pool and executes
them concurrently by spatially partitioning a shared QPU.
Although circuits within a bundle remain logically independent, concurrent
execution can substantially improve hardware utilization.

\vspace{0.02in}
\noindent\textbf{Computation noise.}
Quantum circuits execute on Noisy Intermediate-Scale Quantum (NISQ)
devices~\cite{preskill2018quantum}, which are inherently susceptible to noise
from quantum gates, measurements, and decoherence.
Our measurements further show that these error rates vary across QPUs;
detailed backend statistics are provided in
Appendix~\ref{sec:ibm_logging}.
To improve reliability, \emph{error mitigation} techniques transform noisy
circuits into executions that are less susceptible to hardware errors
~\cite{ayanzadeh2023frozenqubits,tang2021cutqc,giortamis2025qos}.
For example, circuit-cutting-based methods decompose a large circuit into
smaller fragments, reducing interference and shortening individual QPU
executions.
Figure~\ref{fig:fidelity_comparison} compares FrozenQubits,
CutQC, and QOS, normalized to Qiskit.
{Boxplots show medians and quartiles, with circles marking outliers.}
QOS achieves the highest fidelity, improving relative fidelity by up to
$347.7\times$, demonstrating the importance of effective error-mitigation
software.

\begin{figure}[t]
    \centering
    \includegraphics[width=1.0\linewidth, trim={0.2cm 0 0.2cm 0}, clip]{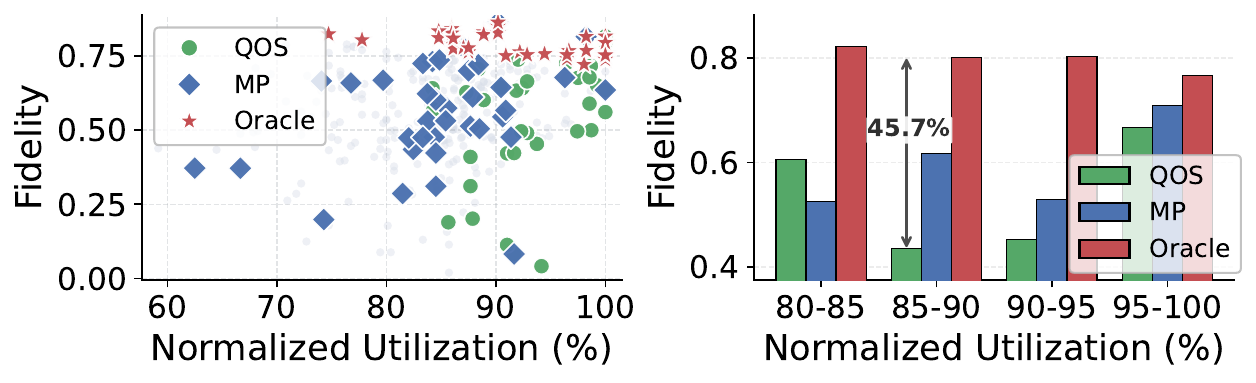}
    \caption{Existing multiprogramming methods achieve poor trade-offs of fidelity and utility.}
    \label{fig:limitation_underutilization}
\end{figure}
\subsection{Pitfalls in Quantum Software Design}

Due to the high design cost, quantum software is often ad-hoc and built around
handcrafted heuristics. 
These heuristics restrict how the large design space is explored and can lead to either suboptimal decisions or excessive search overhead. 

\vspace{0.02in}
\noindent\textbf{Multiprogramming.}
Multiprogramming determines which circuits are bundled for concurrent
execution on a shared QPU~\cite{das2019case,giortamis2025qos}. Existing
methods rank feasible bundles using manually designed scoring functions whose
components and weights approximate the fidelity--utilization trade-off.
Because this fixed scoring rule explores only a restricted set of trade-offs,
the selected bundles can remain far from the offline optimum.

Figure~\ref{fig:limitation_underutilization} compares MP~\cite{das2019case} and QOS~\cite{giortamis2025qos} with an offline-optimal Oracle on IBM\_Torino. 
Normalized utilization characterizes how effectively a bundle uses QPU resources across space and
time~\cite{giortamis2025qos}. 
In Figure~\ref{fig:limitation_underutilization} (left), the top 10\% bundles selected by each method for 24-qubit workloads are highlighted.
Oracle selects bundles near the top-right corner, whereas MP favors central bundles and QOS improves utilization at the cost of fidelity.
Figure~\ref{fig:limitation_underutilization} (right) further reports the average fidelity in each utilization range. At 85\%--90\% utilization, QOS achieves 45.7\%, lower fidelity than Oracle, demonstrating the suboptimality of handcrafted heuristics.

\vspace{0.02in}
\noindent\textbf{Error mitigation.}
{A key design choice in circuit-cutting-based error mitigation
is the maximum fragment size~\cite{mitarai2021constructing,peng2020simulating}.}
%
%
Smaller fragments can improve fidelity, but increase both partitioning cost and the number of QPU executions.
Existing methods therefore sweep candidate fragment sizes until finding one
that satisfies the execution budget. 
Because each candidate requires solving a costly partitioning problem, only a limited portion of the design space can be explored within a practical time budget.

Figure~\ref{fig:cut_pass_timing} shows the circuit-cut search time of gate
virtualization (GV)~\cite{mitarai2021constructing} and wire cutting
(WC)~\cite{peng2020simulating} for QAOA-R3 as the input circuit grows, with
the target fragment size fixed at 7 as in QOS~\cite{giortamis2025qos}. WC
takes 1.11 seconds at 16 qubits, but reaches the 60-second search limit at
18 qubits and remains capped thereafter. GV grows more gradually, but its
search time increases from 0.64 seconds at 20 qubits to 3.40 seconds at
22 qubits and over 10 seconds at 24 qubits. This rapidly increasing cost makes
handcrafted fragment-size exploration inefficient for larger circuits.

\begin{lstlisting}[
  language=Python,
  frame=single,
  basicstyle=\small\ttfamily,
  escapeinside={|}{|},
  caption={LLM-based Evolution Engine},
  label={lst:selfevo},
]
|{\color{dbgreen}codebase}|.add(|$\theta_0$|, |$r_0$|)
while not converged:
    |$\Theta$| = |{\color{dbgreen}codebase}|.sample()
    prompt = |{\color{promptblue}prompter}|.build(|$\Theta$|)
    |$\theta$| = |{\color{llmred}llm}|.generate(prompt)
    |r| = |{\color{evalorange}verifier}|.execute(|$\theta$|)
    |{\color{dbgreen}codebase}|.add(|$\theta$|, |r|)
\end{lstlisting}

\begin{figure*}[t]
    \centering
    \includegraphics[width=\linewidth]{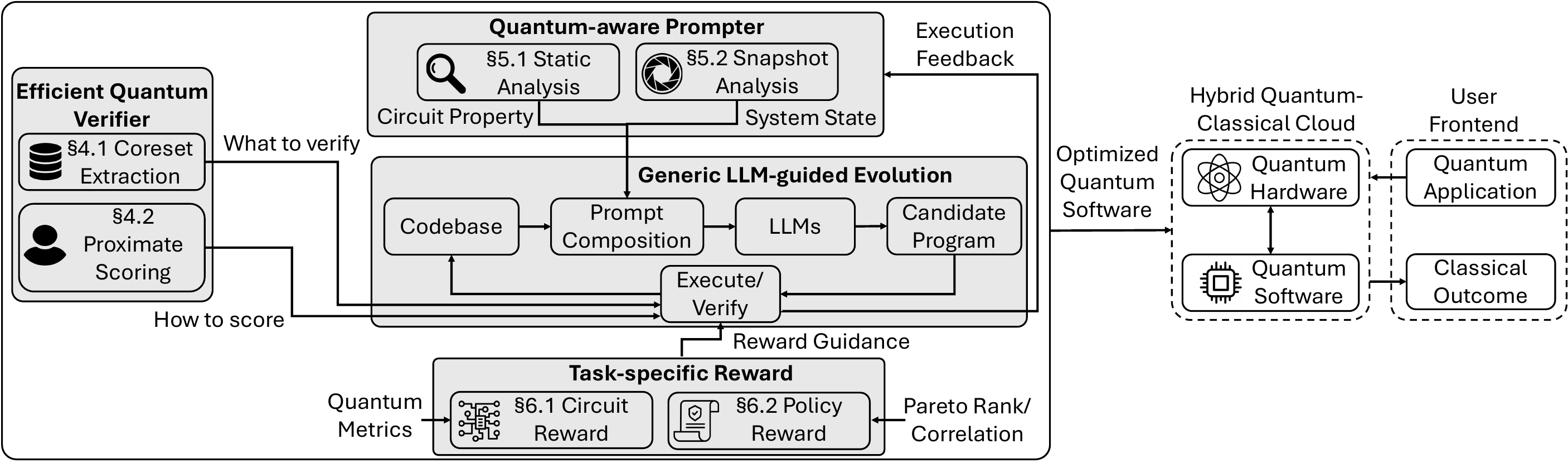}
    \caption{\name's quantum-aware harness around generic LLM-guided evolution, integrating efficient verification, quantum-aware prompting, and task-specific rewards into a closed evolution loop.}
    \label{fig:system-workflow}
\end{figure*}

\subsection{LLM-Guided Evolutionary Search}

Recent LLM-guided evolutionary frameworks use LLMs as mutation operators to
iteratively generate and refine program code~\cite{openevolve,novikov2025alphaevolve}.
As shown in Listing~\ref{lst:selfevo}, a generic evolution loop consists of a
codebase, a prompter, an LLM, and a verifier.
Starting from an initial program $\theta_0$, each iteration samples prior
programs from the codebase, constructs an evolution prompt, generates a
candidate program $\theta$, and executes it to obtain reward $r$, which is
added back to the codebase to guide subsequent search.

\vspace{0.02in}
\noindent\textbf{Preliminary Study.}
We first apply this generic workflow to error-mitigation software using
OpenEvolve~\cite{openevolve} and Gemini-3-Flash~\cite{gemini3flash2025}.
Across 100 iterations, evolution discovers three reward-improving programs,
reducing runtime by up to 19.9\%, demonstrating the potential of LLM-guided
evolution for quantum software.
However, directly applying the generic loop exposes three limitations discussed
in \cref{sec:introduction}: verifying each mutation requires costly quantum
simulation or execution, the LLM receives little quantum-specific context
beyond final rewards, and different quantum software components require
different reward formulations.
These limitations motivate \name's designs introduced next.

\section{System Design}
\label{sec:sys_desgin}

Figure~\ref{fig:system-workflow} illustrates how \name augments a generic
LLM-guided evolutionary search agent and interfaces with a hybrid
quantum--classical cloud.
The \emph{quantum-aware harness} provides domain-specific mechanisms for
verification, prompting, and reward design, while the underlying evolution
agent performs generic program selection and mutation.
The optimized quantum software is then 
integrated into the quantum software stack
to control
quantum execution and return classical outcomes to user applications.

Within the evolution loop, the harness provides three forms of quantum-specific
support.
First, the verifier determines \emph{what to verify and how to score it}:
coreset extraction reduces the circuits used for verification
(\cref{sec:coreset_extraction}), while proxy scoring replaces expensive
fidelity evaluation with lightweight proxy scores derived from circuit
properties (\cref{sec:approx_scoring}).
Second, the prompter determines \emph{what context to expose}: static analysis
summarizes circuit properties (\cref{sec:static_analysis}), while snapshot
analysis captures intermediate system states during execution
(\cref{sec:snapshot_analysis}).
Third, task-specific rewards determine \emph{how to evaluate} heterogeneous
quantum programs, using quantum metrics for circuit optimization
(\cref{sec:circuit_level_reward}) and policy-specific rewards for runtime
policies (\cref{sec:policy_reward}).
%
Together, these components form a closed feedback loop: the verifier provides
low-cost measurements, the prompter exposes quantum-specific execution context, and the rewards convert 
{evaluation results}
into guidance for subsequent mutations.

\begin{figure}[t]
  \begin{minipage}[t]{0.48\linewidth}
    \centering
    \includegraphics[width=\linewidth]{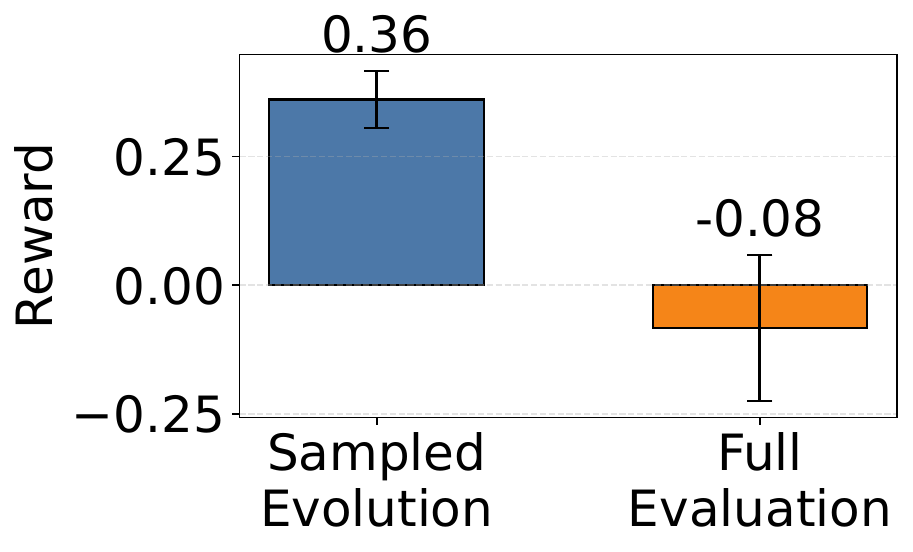}
    \caption{Random selection performance.}
    \label{fig:randq_selection}
  \end{minipage}
  \hfill
  \begin{minipage}[t]{0.46\linewidth}
    \centering
    \includegraphics[width=\linewidth]{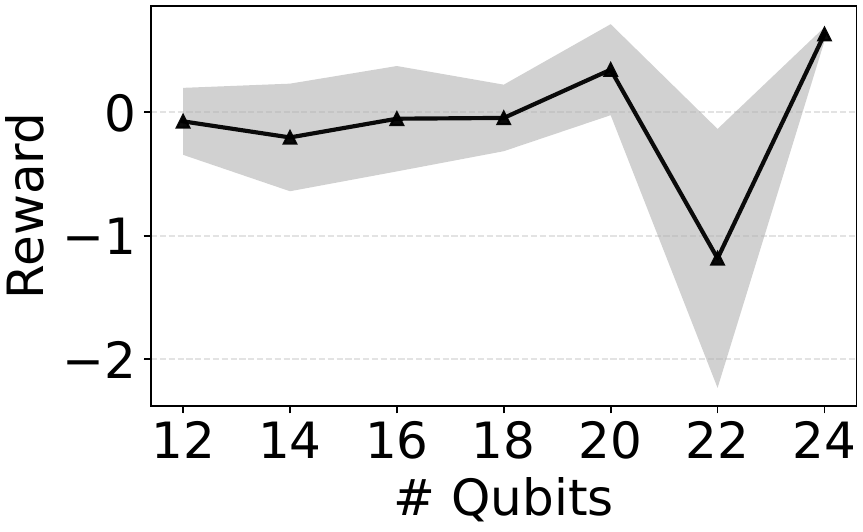}
    \caption{Post-evolution reward across circuit scales.}
    \label{fig:gem3flash_qubit_sampling_randq_reward_vs_qubits}
  \end{minipage}
\end{figure}
\section{Efficient Quantum Verifier}
\label{sec:verifier}
We reduce quantum verification cost along two dimensions: selecting a compact
verification set through evolution-hardness-guided coreset extraction, and
replacing expensive fidelity evaluation with lightweight approximate scoring.

\subsection{Evolution-Hardness-Guided Coreset Extraction}
\label{sec:coreset_extraction}

Ideally, every candidate program should be verified on the complete test
suite. However, full-suite verification is prohibitively expensive and
substantially slows evolution. We therefore use a compact
\emph{verification set} to score candidate programs during evolution, and
evaluate only the final evolved program on the complete suite.

A natural baseline is to uniformly sample circuits from the full suite.
However, such a sample may underrepresent cases that remain difficult for evolution.
To illustrate, we conduct an error-mitigation experiment in which uniformly
sampled circuits are used throughout evolution with
Gemini-3-Flash~\cite{gemini3flash2025}. We defer the reward design to
\cref{sec:reward_design}. As shown in
Figure~\ref{fig:randq_selection}, the final program achieves a reward of
$0.36$ on the sampled verification set, but only $-0.08$ on the complete test
suite. This gap indicates that evolution can overfit to the sampled circuits.





Our key insight is to use the behavior of evolution itself to identify
challenging circuit scales.
Specifically, we uniformly sample circuits from the full suite and use the
sampled set throughout an evolution run.
We then evaluate the evolved program separately at each qubit count.
We repeat this process across random samples to reduce sampling variance.
We call the resulting scale-wise difficulty \emph{evolutionary hardness}: a
lower post-evolution reward indicates that uniformly guided evolution performs
more poorly at that scale, making it a more challenging target for verification.

Formally, let $\mathcal{S}^{(k)}$ denote the $k$-th uniformly sampled
verification set, and let $R_q(\theta)$ denote the reward of program $\theta$
on circuits with qubit count $q$.
We select
\begin{equation}
    q^*
    =
    \arg\min_q
    \frac{1}{K}
    \sum_{k=1}^{K}
    R_q\left(
        \operatorname{Evolve}(\mathcal{S}^{(k)})
    \right),
\end{equation}
where $K$ is the number of repeated random samples.
As shown in
Figure~\ref{fig:gem3flash_qubit_sampling_randq_reward_vs_qubits},
the evolved programs achieve the lowest reward on $22$-qubit circuits,
identifying this scale as the most evolutionarily challenging.

We then construct the final coreset by sampling from all available circuit types at the
selected qubit count. 
Unlike selecting a few globally hardest circuits, this group-level design preserves circuit diversity while focusing verification on a challenging scale. 
%
As shown in Figure~\ref{fig:best_overall_across_sizes}, focusing evolution on
this hard scale does not sacrifice performance across other circuit types or
qubit counts.

\subsection{Evolution-Oriented Approximate Scoring}
\label{sec:approx_scoring}

\begin{figure}[t]
  \begin{minipage}[!htb]{0.49\linewidth}
    \centering
    \includegraphics[width=\linewidth]{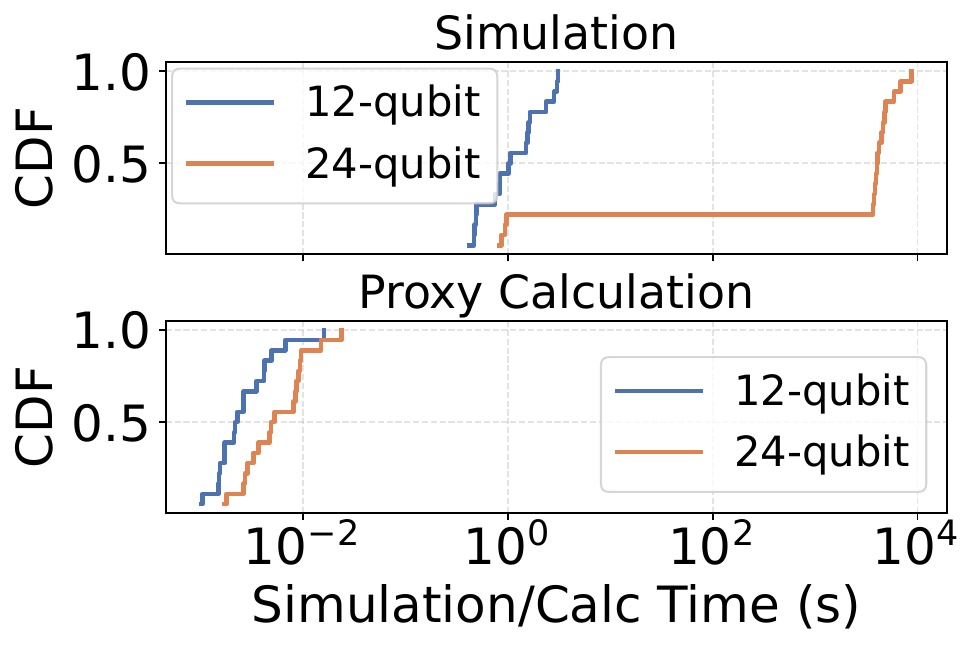}
    \caption{Simulation using Qiskit.}
    \label{fig:qiskit_nocut_simulation_cdf}
  \end{minipage}
  \hfill
  \begin{minipage}[!htb]{0.49\linewidth}
    \centering
    \includegraphics[width=\linewidth]{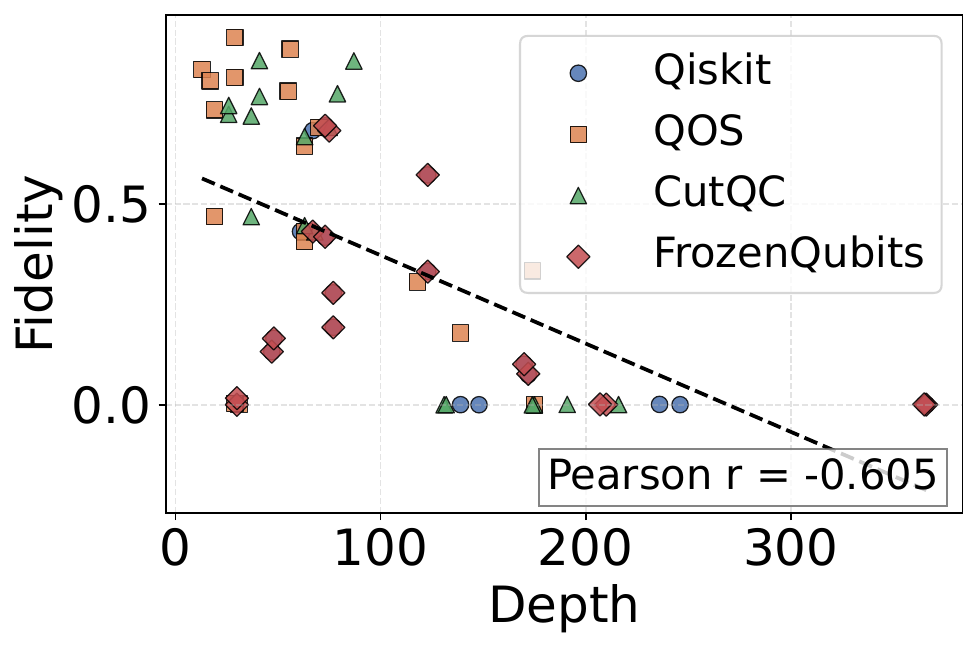}
    \caption{Correlation of depth and fidelity.}
    \label{fig:proxy_fidelity_corr}
  \end{minipage}
\end{figure}

Coreset extraction reduces the number of verification cases, but evaluating
each case remains expensive when fidelity must be obtained through simulation
or real QPU execution. As shown in
Figure~\ref{fig:qiskit_nocut_simulation_cdf} (top), simulating a 12-qubit
circuit takes less than 10 seconds, whereas a 24-qubit circuit can be nearly
three orders of magnitude slower. Real QPU execution additionally incurs
monetary cost and access delays.

Our key insight is that evolution does not require exact fidelity for every
candidate program. It instead needs a low-cost signal that distinguishes
promising candidates and guides the search. 
We therefore replace exact fidelity evaluation during evolution with lightweight proxies derived from circuit properties.
For individual circuits, we use depth and CNOT
count: depth captures the critical-path length, while CNOT count reflects the
use of error-prone two-qubit gates. Prior work has shown both to be useful
fidelity indicators~\cite{hopf2025improving,ayanzadeh2023frozenqubits,
wilson2021empirical}. Figure~\ref{fig:proxy_fidelity_corr} further shows a
moderate negative correlation between depth and fidelity ($-0.605$). Although
these proxies do not predict fidelity exactly, they are inexpensive to compute
(Figure~\ref{fig:qiskit_nocut_simulation_cdf}, bottom) and provide effective
feedback for evolution.

For bundles that execute multiple circuits concurrently, absolute depth or
CNOT count alone does not capture how work is distributed between the
sub-circuits. We therefore use their \emph{relative balance} as a
bundle-level proxy. For a circuit property $v$ (\eg depth or CNOT count), we
define

\begin{equation}
\label{eq:relative_balance}
    r(v_1,v_2)
    =
    {\min\{v_1,v_2\}}/
         {\max\{v_1,v_2,1\}},
\end{equation}

where $v_1$ and $v_2$ are the property values of the two bundled circuits.
The score lies in $[0,1]$, with a smaller value 
indicating greater imbalance.
These lightweight, bundle-aware proxies replace
expensive fidelity evaluation and substantially accelerate verification
during evolution.

\section{Quantum-Aware Prompter}
\label{sec:prompter}

The base prompt in OpenEvolve~\cite{openevolve} provides the LLM with program
code and reward feedback, but little information about the quantum workloads
being optimized or how a program changes them during execution. This semantic
gap makes it difficult for the LLM to diagnose failures and generate
meaningful mutations. Our key insight is to expose both \emph{what} circuits
the program operates on and \emph{how} their states evolve during execution.
We therefore augment the prompt with two complementary forms of quantum-aware
context: static circuit profiles and execution snapshots.

\subsection{Quantum Circuit Static Analysis}
\label{sec:static_analysis}

Circuit layouts alone provide limited information about their execution
characteristics. We therefore statically analyze each circuit and summarize
it as a compact metadata profile. The profile combines basic statistics from
Qiskit~\cite{gadi_aleksandrowicz_2019_2562111} with DAG-derived features from
SupermarQ~\cite{tomesh2022supermarq}, including circuit width, depth, and CNOT
count. Width captures the number of required qubits, depth approximates the
critical execution path, and CNOT count reflects the use of noise-sensitive
two-qubit operations~\cite{barenco1995elementary}. For example, the GHZ
circuit in Figure~\ref{fig:circuit_example} has width 3, depth 3, and two CNOT
gates.

These features provide the LLM with a concise description of circuit scale,
structure, and noise sensitivity. By relating the metadata to program rewards
and execution behavior, the LLM can identify workload-dependent patterns and
develop more effective optimization heuristics. Appendix~\ref{appendix:static_feature_details} lists the complete feature set.

\subsection{Quantum Software Snapshot Analysis}
\label{sec:snapshot_analysis}

Static profiles describe the input circuits, but do not reveal how an evolved
program transforms them. 
A final reward only indicates whether a program performs well but not where execution begins to degrade or which operation causes the change.

\begin{figure}[t]
    \centering
    \includegraphics[width=1\linewidth]{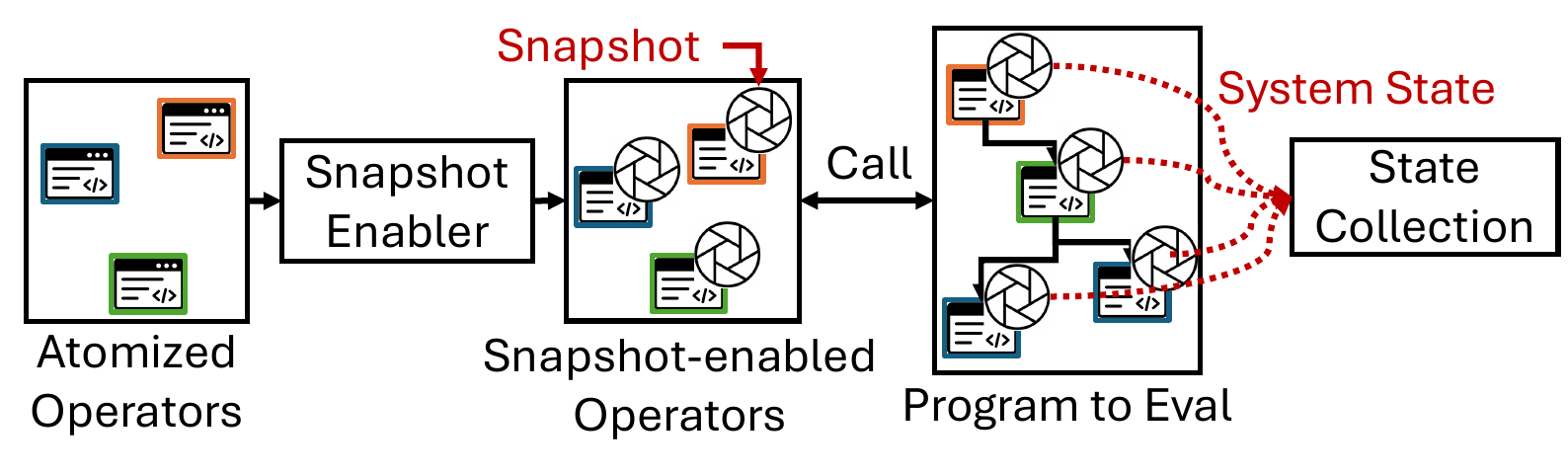}
    \caption{Snapshot Analysis}
    \label{fig:snapshot_analysis}
\end{figure}

We address this limitation through \emph{snapshot analysis}. As illustrated in Figure~\ref{fig:snapshot_analysis}, we instrument key operators that modify quantum-related metrics and record the program state after each operator completes. 
When an evolved program invokes these operators, the collected snapshots expose intermediate circuit properties and execution outcomes at multiple points in the workflow.

This step-by-step context allows the LLM to associate reward changes with specific program decisions, localize ineffective transformations, and generate targeted mutations.

\section{Task-Specific Reward Design}
\label{sec:reward_design}
We select the reward formulation according to how the target program interacts
with quantum circuits.

\subsection{Circuit-level Reward}
\label{sec:circuit_level_reward}
The circuit-level reward evaluates programs that directly modify quantum circuits
(\eg error mitigation and compilation).
Let
$M(j) = \big(m_1(j), m_2(j), \ldots, m_n(j)\big)$
denote the measured system metrics of job $j$, such as fidelity, runtime, or cost.
Because these metrics have different scales and optimization directions, we first
normalize each metric into a dimensionless utility $\tilde{m}_i(j)$, where a larger
value consistently indicates better performance.
We then aggregate the normalized utilities using a weighted reward:
\begin{equation}
\label{eq:weighted_score}
r_{\text{circuit}}(j)
=
\sum_{i=1}^{n} w_i \, \tilde{m}_i(j),
\end{equation}
where $w_i \geq 0$ specifies the relative importance of metric $i$ and
$\sum_i w_i = 1$.
%

\subsection{Policy Reward}
\label{sec:policy_reward}
The policy reward evaluates programs that optimize how circuits are executed on hardware (\eg multiprogramming and scheduling).
We design the policy reward using two components: \emph{Pareto rank}, which 
captures Pareto optimality of conflicting objectives, and \emph{policy correlation}, which 
encourages
generalization of the policy across system settings.

\noindent\textbf{\emph{Pareto Rank.}}
The Pareto optimality ensures the optimality across different trade-offs among conflicting objectives. 
In multiprogramming, for example, fidelity and utilization may be weighted differently depending on system load: under overload, the scheduler may prioritize utilization to sustain throughput, whereas under light load, it may favor fidelity to improve user experience. A fixed weighted score is often too rigid to capture such varying preferences.

We therefore evaluate each candidate job using Pareto dominance. Let $\mathcal{J}$ denote the set of candidate jobs, and let
$M(j) = \big(m_1(j), m_2(j), \ldots, m_n(j)\big)$
denote the metric vector of job $j \in \mathcal{J}$ (higher the better). 
Job $j^\ast$ Pareto dominates job $j$ if it is no worse in every objective and strictly better in at least one (\ie $\forall i \in [1,n], \; m_i(j^\ast) \ge m_i(j)
\quad \text{and} \quad
\exists k \in [1,n], \; m_k(j^\ast) > m_k(j)$).
Based on this relation, we compute Pareto fronts by iterative non-dominated sorting. 

The $t$-th front is defined as
$\mathcal{F}_t
=
\mathrm{Front}\!\left(
\mathcal{J}\setminus \bigcup_{i=1}^{t-1}\mathcal{F}_i
\right),$
where $\mathcal{F}_1$ contains all non-dominated jobs, $\mathcal{F}_2$ contains the non-dominated jobs after removing $\mathcal{F}_1$, and so on. We then define the Pareto-rank reward of job $j$ as
\begin{equation}
\label{eq:pareto_rank}
r_{\text{pareto}}(j)=t, \quad \text{if } j\in\mathcal{F}_t.
\end{equation}
Lower rank is better. By favoring jobs with smaller Pareto rank, the evolved policy is encouraged to remain close to the non-dominated frontier.

\vspace{0.02in}
\noindent\textbf{\emph{Policy Correlation.}}
Pareto optimality alone is insufficient: the policy should also generalize across scheduling settings.
In practice, a policy selects only a subset of jobs from a candidate pool, so rewarding only the final outcome can overfit the policy to a particular workload composition.
Our key insight is to evaluate how well the policy's decisions align with the ground truth.
Let $p(j)$ denote the measure assigned by the policy to job $j$ (\eg its predicted Pareto rank), and let $g(j)$ denote the corresponding ground-truth measure (\eg its offline-optimal rank).
We define the policy-correlation reward as
\begin{equation}
\label{eq:policy_corr}
r_{\text{corr}}
=
\mathrm{corr}\big(p(j), g(j)\big),
\end{equation}
where $\mathrm{corr}(\cdot,\cdot)$ is the Pearson correlation over all candidate jobs in $\mathcal{J}$.
A correlation close to $1$ indicates that the policy closely reproduces the ground-truth ordering.
Unlike outcome-based rewards computed only over selected jobs, this provides a denser signal that encourages generalizable policy decisions.

Finally, we combine Pareto rank and policy correlation:
\begin{equation}
\label{eq:policy_reward}
r_{\text{policy}}
=
\alpha \frac{1}{\bar{r}_{\text{pareto}}}
+
(1-\alpha) r_{\text{corr}},
\end{equation}
where $\bar{r}_{\text{pareto}}$ is the average Pareto rank of the selected jobs and $\alpha\in[0,1]$ controls the relative importance of Pareto optimality and generalization.
We use $1/\bar{r}_{\text{pareto}}\in(0,1]$ to convert the lower-is-better Pareto rank into a higher-is-better reward term.
The resulting reward favors policies that select jobs near the Pareto frontier while preserving decision orderings 
intended to
generalize across workload conditions.

\section{Implementation}
\noindent\textbf{Evolution Engine.}
We run the evolution loop on a Linux server with 32-core AMD Ryzen~9 CPU @ 4.95\,GHz based on OpenEvolve~\cite{openevolve}.
LLM inference is performed via external API calls, and verification 
runs single-threaded.
To ensure system efficiency in the error mitigation, we enforce a 60-second execution deadline per circuit.
%
%
{We conduct three independent 100-iteration trials per setting. The reported performance is averaged across the three runs, and the error bars represent $\pm$ one standard deviation.}


\vspace{0.02in}
\noindent\textbf{Reward.}
We tailor the reward function to each task's requirements.
For multiprogramming, we use the policy reward
(Equation~\ref{eq:policy_reward}), averaging the Pareto-rank term over the
top 1\% candidate bundles.
We set $\alpha=0.5$ and use the depth ratio as a lightweight fidelity proxy
when computing Pareto ranks.
For error mitigation, we instantiate the circuit-level reward
(Equation~\ref{eq:weighted_score}) using depth, CNOT count, and runtime.
Specifically, $\tilde m_d$, $\tilde m_c$, and $\tilde m_t$ denote the
normalized depth, CNOT-count, and runtime metrics, respectively, each averaged
across benchmarks and direction-aligned so that larger values are better.
We use relative weights $1:1:\frac{1}{4}$ for
$\tilde m_d:\tilde m_c:\tilde m_t$, prioritizing circuit-complexity reduction
while still rewarding lower runtime.

\vspace{0.02in}
\noindent\textbf{Prompter.}
To profile program behavior, we extract static features from Qiskit~\cite{gadi_aleksandrowicz_2019_2562111} and Supermarq~\cite{tomesh2022supermarq}, encompassing eight structural metrics (\eg depth, gate counts) and six communication-centric properties (\eg entanglement ratio, parallelism).
The snapshot analysis includes 1) the cost search trace for 22-qubit circuits in error mitigation and 2) the Pareto rank and system metrics of top 1\% circuit bundles from the pool for multiprogramming.
See Appendix~\ref{appendix:configurations} for complete prompt examples.

\section{Evaluation}
Our evaluation answers the following research questions:
\begin{enumerate}[leftmargin=*]
    \item How effectively can \name automate quantum software design in system metrics? (\cref{sec:overall_performance})
    \item How robust is \name? (\cref{sec:generalization})
    \item Where does QSA's performance gain come from? (\cref{sec:performance_breakdown})
    \item How does each harness component contribute? (\cref{sec:ablation_study})
\end{enumerate}

\begin{figure*}
    \centering
    \begin{minipage}[t]{0.48\linewidth}
        \centering
        \includegraphics[width=\linewidth]{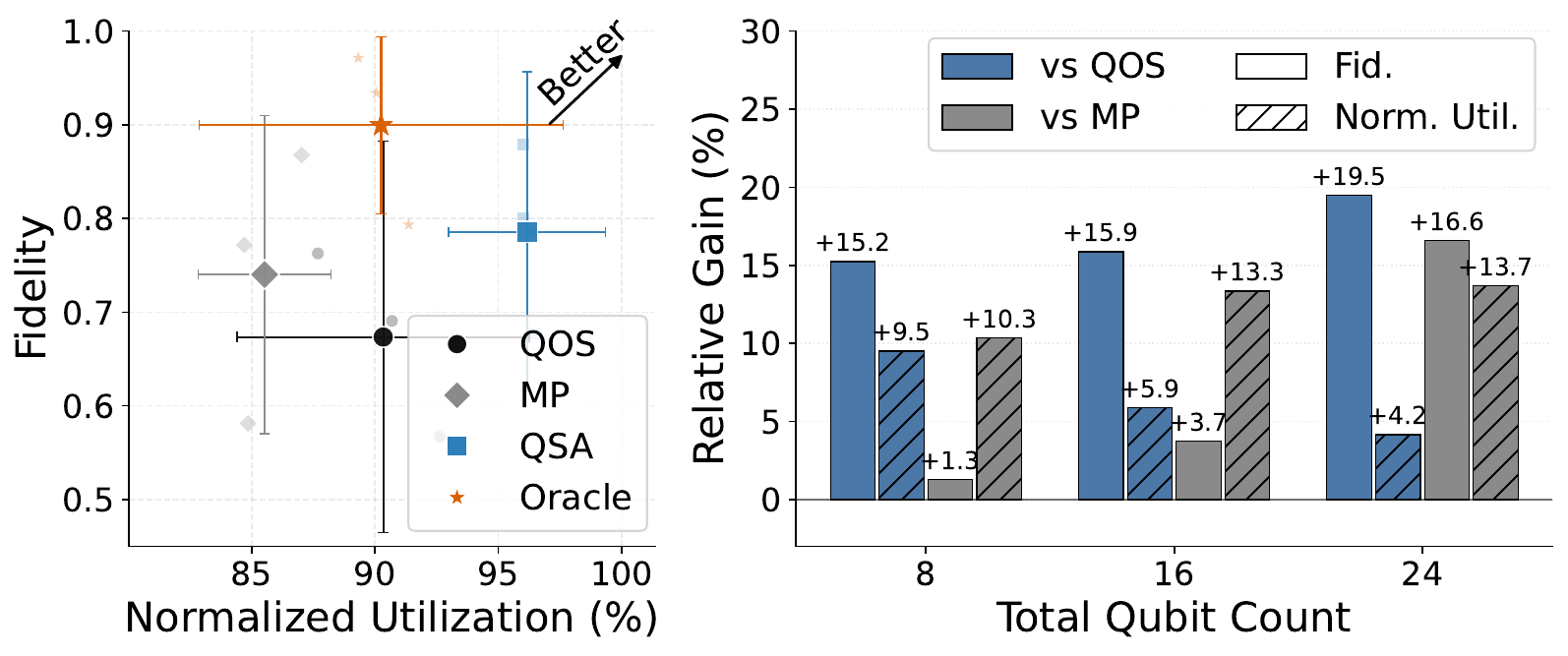}
        \caption{Multiprogramming result.}
        \label{fig:multiprogramming_result}
    \end{minipage}
    \hfill
    \begin{minipage}[t]{0.48\linewidth}
        \centering
        \includegraphics[width=\linewidth]{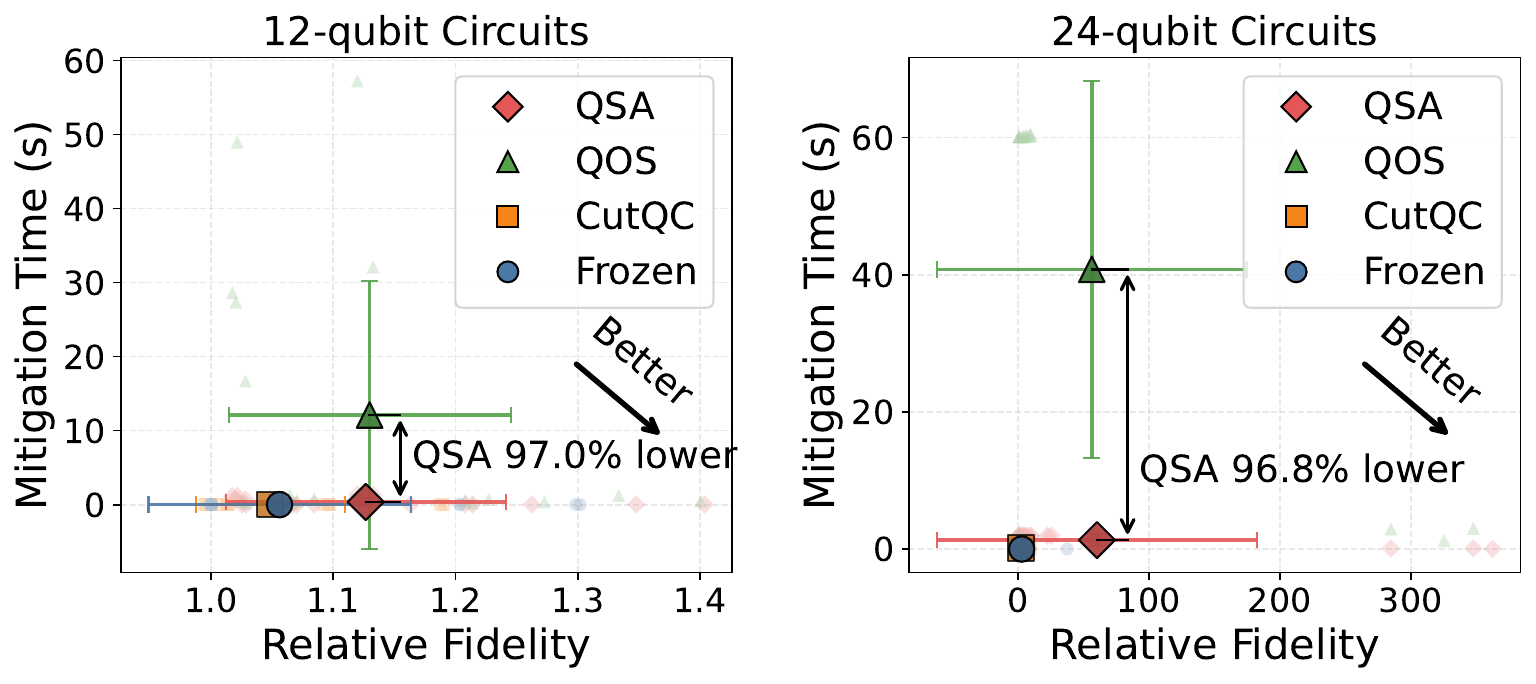}
        \caption{Error mitigation result.}
        \label{fig:error_mitigation_result}
    \end{minipage}
\end{figure*}
\begin{figure}[h]
    \centering
    \begin{subfigure}[t]{0.49\linewidth}
        \centering
        \includegraphics[width=\linewidth]{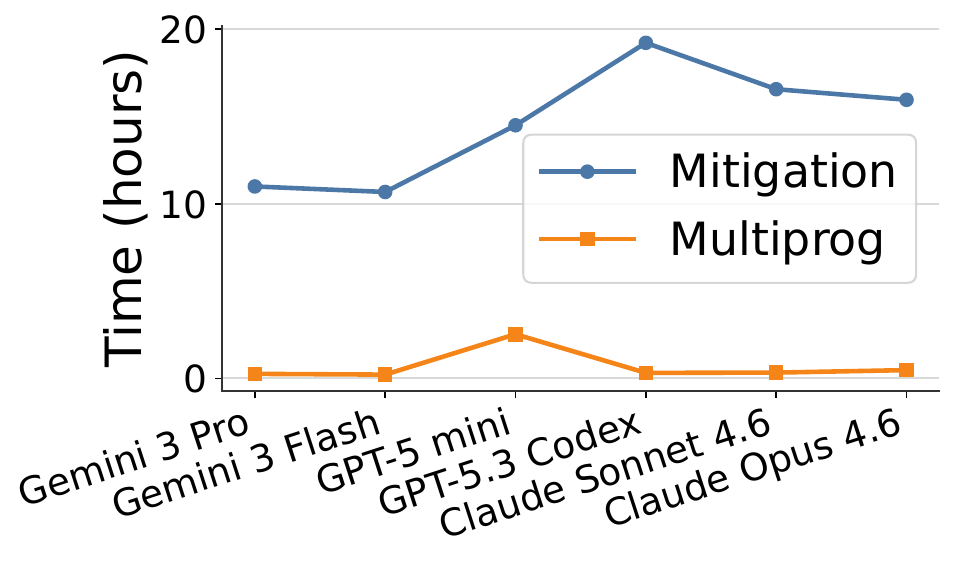}
        \caption{Time Cost.}
        \label{fig:mp_runtime}
    \end{subfigure}
    \hfill
    \begin{subfigure}[t]{0.49\linewidth}
        \centering
        \includegraphics[width=\linewidth]{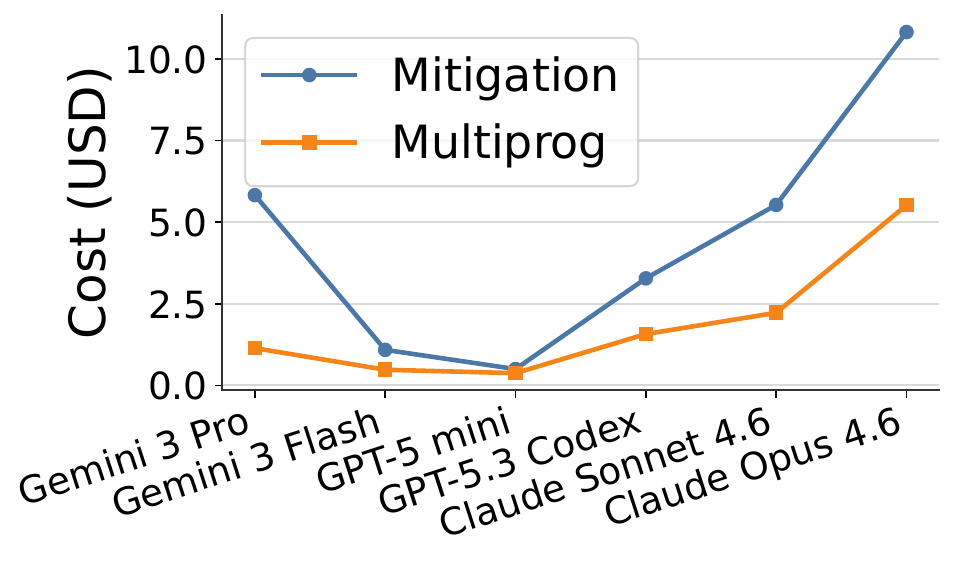}
        \caption{Monetary Cost.}
        \label{fig:mp_cost}
    \end{subfigure}
    \caption{Time and monetary cost across LLM APIs.}
    \label{fig:em_mp_cost}
\end{figure}
\subsection{Experimental Setup}
\label{sec:exp_setup}
\noindent\textbf{Quantum Hardware.}
Experiments are conducted on IBM Heron processors~\cite{ibm_quantum_platform} to evaluate circuit performance.
We evaluate \name on the 133-qubit Torino and 156-qubit Marrakesh QPUs using 8,192 measurement shots per execution.

\vspace{0.02in}
\noindent\textbf{Metrics.}
We evaluate the following metrics:
(1) {\em Hellinger Fidelity:} A score from 0 to 1 measuring the similarity between experimental results and the noise-free ideal (Equation~\ref{sec:fidelity_definition}). 
{\em Relative fidelity} represents the raw fidelity normalized by Qiskit results.
(2) {\em Normalized Effective Utilization:} Following QOS~\cite{giortamis2025qos}, normalized utilization is the combined spatial and temporal utilization of a bundled execution.
It accounts for how many qubits are occupied and how long they are occupied when bundled circuits have different runtimes.
%
For brevity, we refer to this metric as normalized utilization.
(3) {\em Mitigation Time:} The latency incurred during the pre-processing of large circuits into smaller ones for improving fidelity.
(4) {\em Quantum Overhead:} The expansion factor ($\times$) in the total number of circuit fragments.
(5) {\em Static Properties:} The CNOT gate count (CNOT) and circuit depth (Depth), which indicate circuit complexity and expected error rates.

\vspace{0.02in}
\noindent\textbf{Baselines.}
We compare \name to two sets of baselines to comprehensively evaluate its performance.
For multiprogramming, we compare against MP~\cite{das2019case}, which enables
concurrent execution through interference-aware scheduling, and QOS~\cite{giortamis2025qos},
which uses utility/fidelity-aware heuristics to place compatible workloads on
shared QPU resources.
We exclude HyperQ~\cite{tao2025quantum} and QVM~\cite{tornow2025qvm} because they primarily provide
execution abstractions rather than bundle-ranking policies.
For error mitigation, we compare against Qiskit~\cite{gadi_aleksandrowicz_2019_2562111}, CutQC~\cite{tang2021cutqc},
FrozenQubits~\cite{ayanzadeh2023frozenqubits}, and QOS~\cite{giortamis2025qos}, representing compiler optimization,
circuit cutting, qubit freezing, and an integrated mitigation pipeline,
respectively.



\vspace{0.02in}
\noindent\textbf{Workloads.}
Following QOS~\cite{giortamis2025qos}, we evaluate error mitigation using nine representative NISQ algorithms
from the Supermarq~\cite{tomesh2022supermarq}, MQT-Bench~\cite{quetschlich2023mqt}, and QASM-Bench~\cite{li2023qasmbench} suites: GHZ, W-State, Bernstein-Vazirani (BV), Hamiltonian Simulation (HS), Quantum-enhanced Support Vector Machine (QSVM), Two-Local Ansatz (TL), Variational Quantum Eigensolver (VQE-n), and Approximate Optimization Algorithm (QAOA-R/P).
The algorithms' circuits can be scaled by the number of qubits, providing comprehensive coverage of criteria for evaluating contemporary quantum workloads. 

For multiprogramming, we use dual-program workloads that pairs the above heterogeneous circuits across scales, following prior works~\cite{das2019case, giortamis2025qos}. 
We enforce a fixed target utilization when creating the candidate pool for scheduling: we pair circuits such that their combined qubit count remains identical across all policies. 
This ensures a fair evaluation while isolating the impact of qubit numbers.

\vspace{0.02in}
\noindent\textbf{Mutation Operator.}
Since the evolution requires an LLM engine for mutation, we consider a diverse selection of commercial LLM families:
(1)~\GeminiTP~\cite{google2025gemini3announcement}, (2)~\GeminiTF~\cite{gemini3flash2025}, (3)~\GPTF~\cite{openai_gpt5},  (4)~\GPTFC~\cite{openai2026gpt53codex}, (5)~\ClaudeSonnet~\cite{anthropic2025claude45sonnet}, (6)~\ClaudeOpus~\cite{anthropic2025claude45opus}.
Unless otherwise specified, the results are derived from \GeminiTP, which has the best overall performance.

\begin{figure*}[t]
    \centering
    \begin{minipage}[t]{0.245\linewidth}
        \centering
        \includegraphics[width=\linewidth]{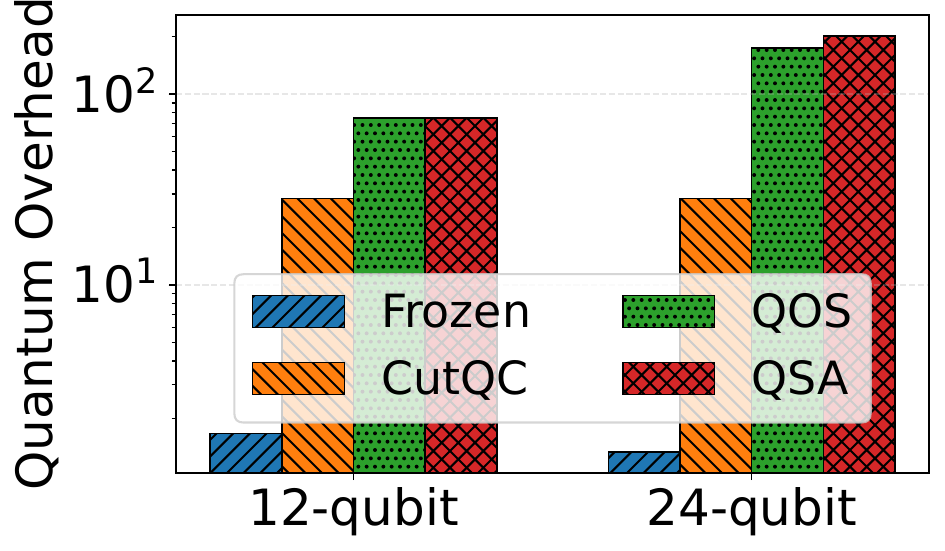}
        \caption{Quantum overhead across circuit scales.}
        \label{fig:panel_avg_jobs_per_bench}
    \end{minipage}
    \hfill
    \begin{minipage}[t]{0.245\linewidth}
        \centering
        \includegraphics[width=\linewidth]{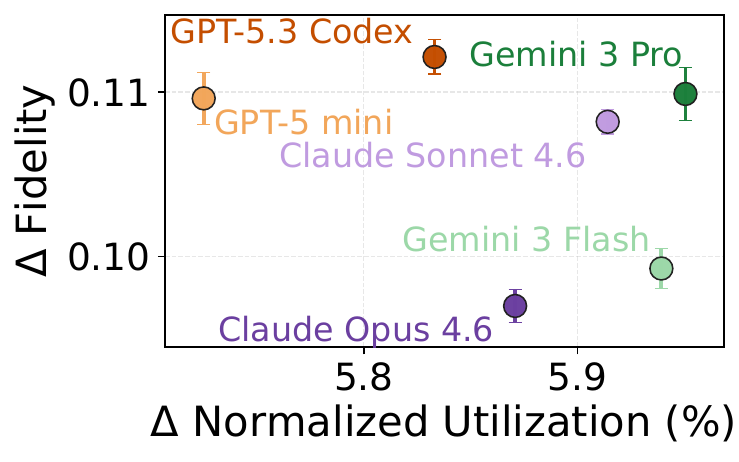}
        \caption{Multiprogramming comparison across LLMs.}
        \label{fig:multiprogramming_model_comparison}
    \end{minipage}
    \hfill
    \begin{minipage}[t]{0.49\linewidth}
        \centering
        \includegraphics[width=\linewidth]{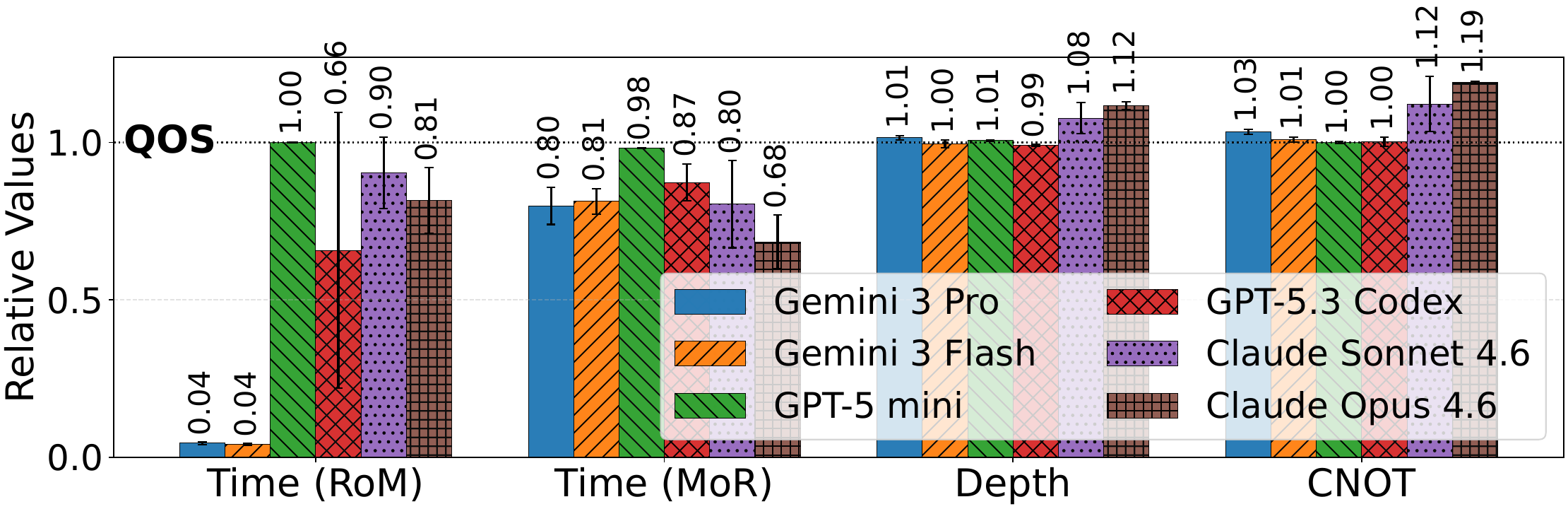}
        \caption{Error mitigation comparison across LLMs.}
        \label{fig:mitigation_model_comparison}
    \end{minipage}
\end{figure*}

\begin{figure*}[t]
    \centering
    \begin{minipage}[t]{0.24\linewidth}
        \centering
        \includegraphics[width=\linewidth]{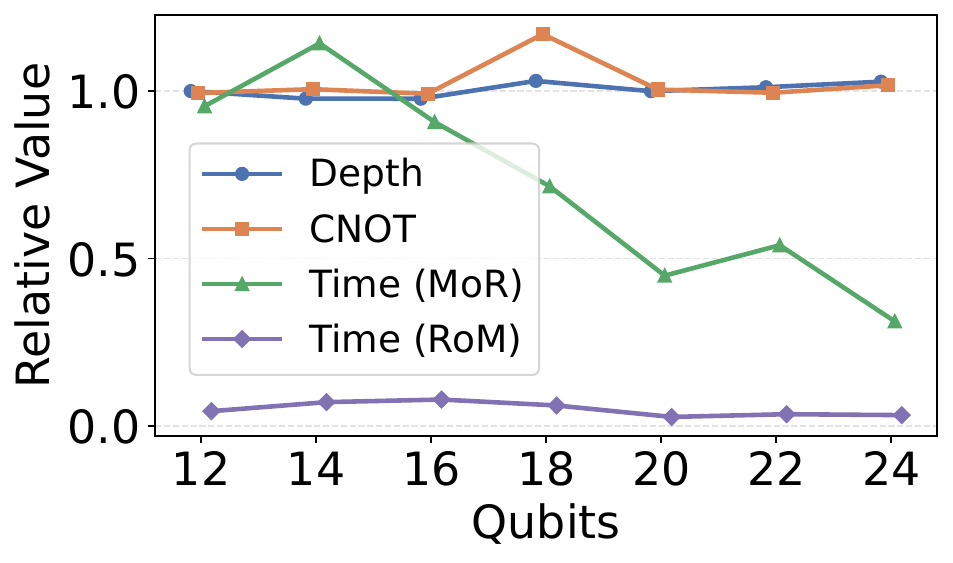}
    \caption{\name performance across circuit scales.}
    \label{fig:best_overall_across_sizes}
    \end{minipage}
    \hfill
    \begin{minipage}[t]{0.24\linewidth}
        \centering
        \includegraphics[width=\linewidth]{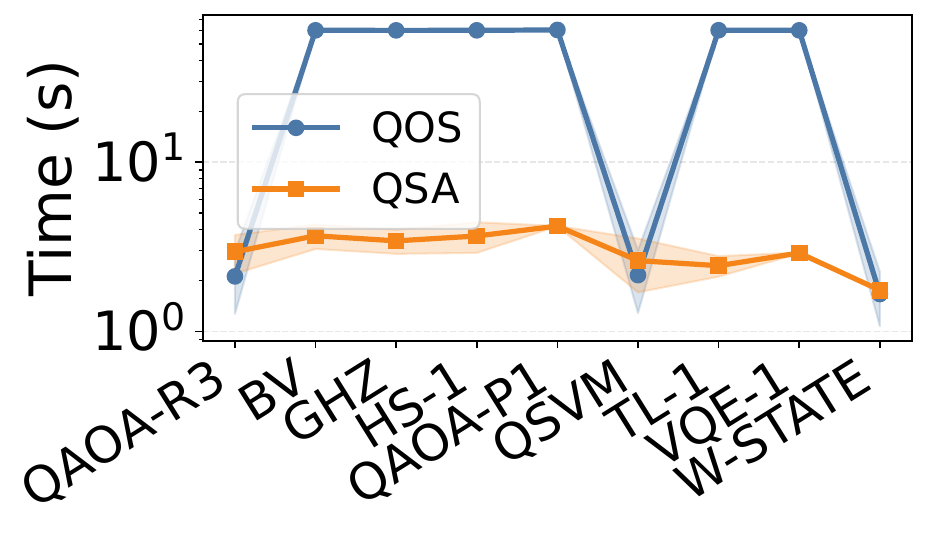}
    \caption{\name performance across circuit topologies.}
    \label{fig:mitigation_stage_breakdown_qos_qose}
    \end{minipage}
    \hfill
    \begin{minipage}[t]{0.24\linewidth}
        \centering
        \includegraphics[width=\linewidth]{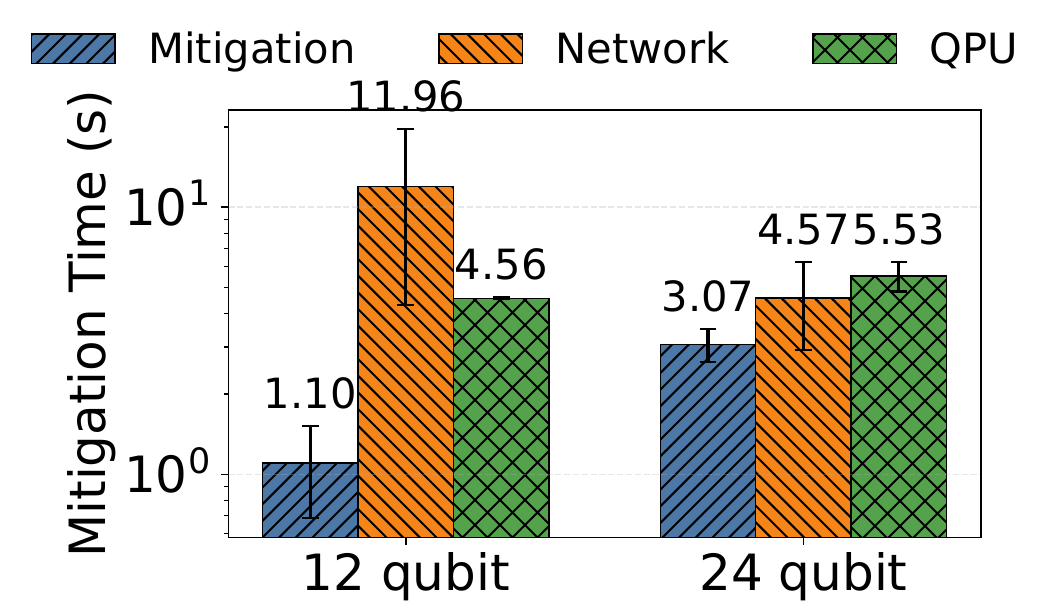}
    \caption{\name time breakdown.}
    \label{fig:qsa_time_breakdown}
    \end{minipage}
    \hfill
    \begin{minipage}[t]{0.24\linewidth}
        \centering
        \includegraphics[width=\linewidth]{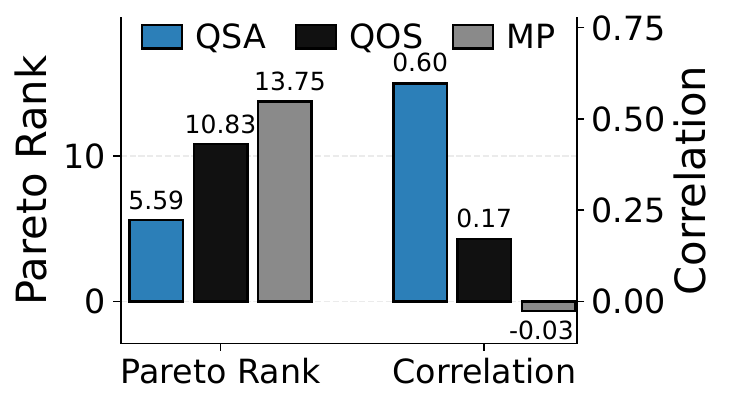}
    \caption{Pareto rank and correlation analysis. }
    \label{fig:figure_qsa_qos_mp_pareto_bar}
    \end{minipage}
\end{figure*}
\subsection{Overall Performance}
\label{sec:overall_performance}
\noindent\textbf{Multiprogramming (Fidelity vs. Utilization).}
We evaluate the efficacy of \name's multiprogramming module using fidelity and normalized utilization.
As shown in Figure~\ref{fig:multiprogramming_result}, \name outperforms both QOS and MP in both fidelity and normalized utilization, while exhibiting a different trade-off from the Oracle, \ie higher utilization but lower fidelity.
The error bars show variance across total qubit count (8, 16, and 24).
Specifically, compared to QOS, \name improves fidelity by 15.2\%--19.5\% and normalized utilization by 4.2\%--9.5\% across different target utilization settings.
Compared to MP, \name improves fidelity by 1.3\%--16.6\% and normalized utilization by 3.7\%--13.7\%.
Overall, the gains become more pronounced when the system aims to use more qubits concurrently.

\vspace{0.02in}
\noindent\textbf{Error Mitigation (Fidelity vs. Mitigation Time).}
We evaluate the error mitigation performance of \name using fidelity and mitigation time.
Figure~\ref{fig:error_mitigation_result} shows the trade-off between mitigation time and relative fidelity for small-scale (12-qubit) and medium-scale (24-qubit) circuits.
\name achieves fidelity comparable to QOS while reducing mitigation time by at least 96.8\%, yielding the best trade-off between fidelity and mitigation time.
We observe a slight fidelity drop on small-scale circuits, but a slight fidelity improvement on medium-scale circuits.
%
These results show that \name maintains strong performance across different
circuit scales and suggest that its benefits may become more pronounced for
larger circuits.

\vspace{0.02in}
\noindent\textbf{Time and Monetary Cost.}
Figure~\ref{fig:em_mp_cost} reports the runtime and monetary cost of evolving error mitigation and multiprogramming across different LLMs.
We estimate monetary cost using the input/output token usage reported by the API together with the official token pricing.
On average, error mitigation evolution costs 10.7--19.2 hours and \$0.5--\$10.8, while multiprogramming costs 0.2--2.5 hours and \$0.4--\$5.5.
Overall, the monetary cost is modest compared to the human effort required to manually design and tune quantum software.
Error mitigation incurs substantially higher runtime because its verification is much more expensive than that of multiprogramming.
%
%
%
Among the evaluated models, \GeminiTF~\cite{gemini3flash2025} and \GPTF~\cite{openai_gpt5} achieve the lowest monetary cost, whereas \ClaudeOpus~\cite{anthropic2025claude45opus} incurs the highest.

\vspace{0.02in}
\noindent\textbf{Quantum Overhead.}
Figure~\ref{fig:panel_avg_jobs_per_bench} shows the quantum overhead of error mitigation.
Across different circuit scales, \name incurs quantum overhead comparable to that of QOS.
Both \name and QOS exhibit higher overhead than CutQC and FrozenQubit because they partition circuits more aggressively.
However, this additional overhead can be amortized through multiprogramming or QPU-level parallelism.

\vspace{0.02in}
Overall, QSA translates automated software design into 4.2--9.5\% higher QPU
utilization and 15.2--19.5\% higher fidelity for multiprogramming, while
removing at least 96.8\% of mitigation preprocessing time 
without sacrificing fidelity, enabling more efficient use of scarce QPU resources.

\subsection{Robustness Across Settings}
\label{sec:generalization}
\noindent\textbf{LLM Models.}
Figure~\ref{fig:multiprogramming_model_comparison} and Figure~\ref{fig:mitigation_model_comparison} show how \name generalizes across modern LLM APIs for both error mitigation and multiprogramming.
For multiprogramming, Figure~\ref{fig:multiprogramming_model_comparison} reports the improvements in fidelity and normalized utilization relative to QOS.
All evaluated LLMs achieve better trade-offs between fidelity and normalized utilization than QOS.
\GeminiTP and \GPTFC form the Pareto frontier, with \ClaudeOpus attaining the worst performance.

For error mitigation, Figure~\ref{fig:mitigation_model_comparison} compares the relative time, depth, and CNOT across different LLM APIs, all normalized to QOS.
For depth and CNOT, we report the circuit-wise mean ratio to QOS.
For time, we report both the \emph{mean of ratios} (MoR) and the \emph{ratio of means} (RoM), with the horizontal line indicating QOS.
Overall, most LLMs substantially reduce mitigation time, with \GeminiTP and \GeminiTF lowering RoM to 0.04 (\ie a 96\% reduction).
The main exception is \GPTF, which fails to improve the reward during evolution.
For depth and CNOT, Gemini and GPT models keep the relative change within 3\%, whereas Claude increases them by 8\%--19\%, suggesting slight overfitting in this setting.
Overall, \name generalizes well across modern LLM APIs, except that \GPTF fails to improve mitigation time and Claude performs worse on circuit properties.

\vspace{0.02in}
\noindent\textbf{Circuit Scales.}
Figure~\ref{fig:best_overall_across_sizes} evaluates \name on quantum circuits ranging from 12 to 24 qubits, with all metrics normalized to those of QOS.
Across circuit sizes, \name preserves circuit quality comparable to QOS: the relative depth and CNOT count remain within 3\% of QOS in nearly all cases, except for the CNOT count of the 18-qubit circuit.
The error-mitigation time under RoM remains consistently below 10\% of that of QOS across all circuit sizes.
Under MoR, the relative error-mitigation time generally decreases as the circuit size increases and exceeds that of QOS only for the 14-qubit circuit.
%
Overall, \name maintains consistent performance across circuit scales,
preserving circuit quality while substantially reducing error-mitigation time.
Moreover, its proxy-based verifier avoids the costly simulations required by QOS, thereby reducing verification overhead.
%

\vspace{0.02in}
\noindent\textbf{Circuit Topologies.}
Figure~\ref{fig:mitigation_stage_breakdown_qos_qose} shows the mitigation time distribution across different topologies of 24-qubit quantum circuits.
QOS fails to maintain low latency for 6 out 9 circuit topologies (\eg BV, GHZ).
While the error mitigation time of \name is slightly higher than that of QOS for QAOA-R3 and QSVM circuits, \name remains a consistent and low error mitigation time.

\begin{figure*}[t]
    \centering
    \begin{minipage}[t]{0.24\linewidth}
        \centering
        \includegraphics[width=\linewidth]{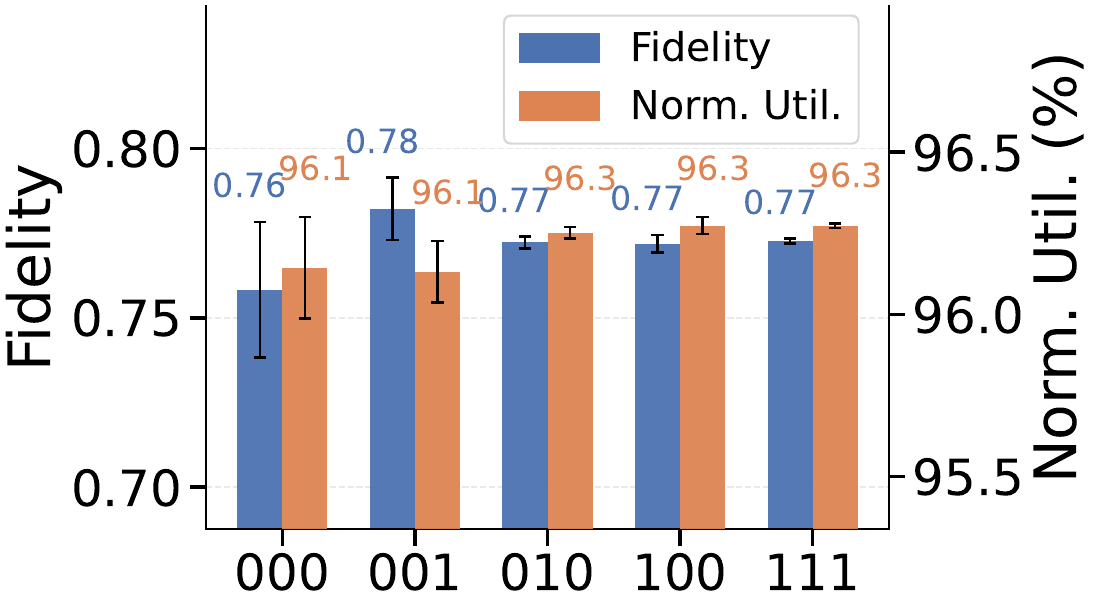}
    \caption{Impact of prompt composition.}
    \label{fig:art_ablation}
    \end{minipage}
    \hfill
    \begin{minipage}[t]{0.24\linewidth}
        \centering
        \includegraphics[width=\linewidth]{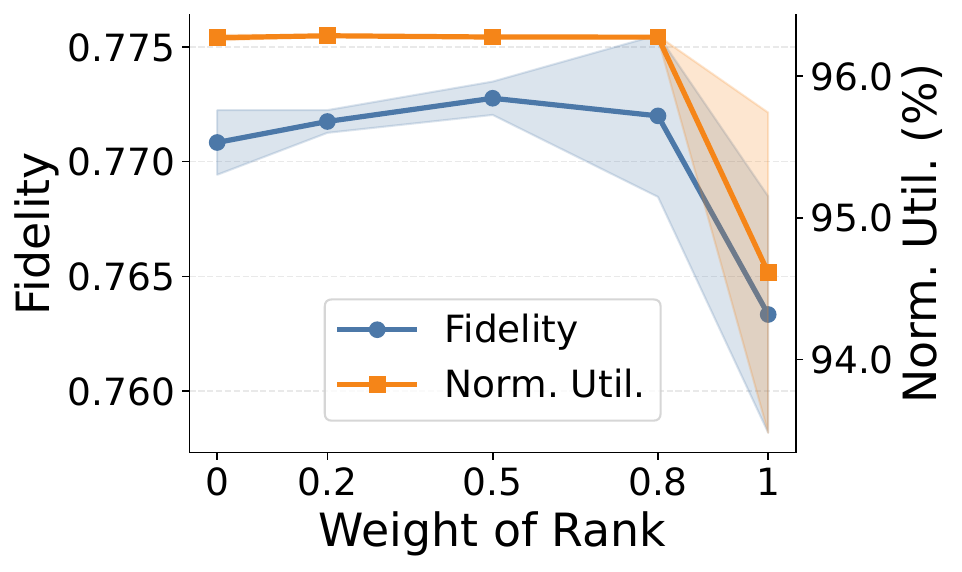}
    \caption{Impact of score weights.}
    \label{fig:score_ablation}
    \end{minipage}
    \hfill
    \begin{minipage}[t]{0.24\linewidth}
        \centering
        \includegraphics[width=\linewidth]{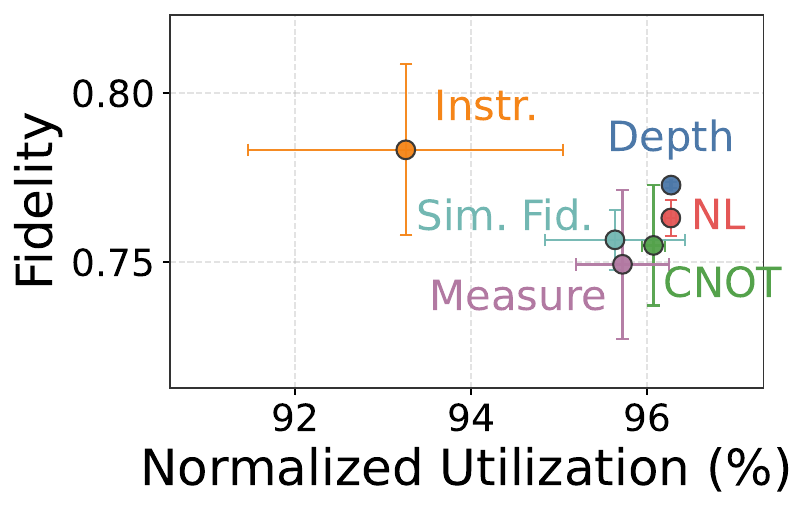}
    \caption{Impact of proxy on fidelity and utilization.}
    \label{fig:impact_of_proxy}
    \end{minipage}
    \hfill
    \begin{minipage}[t]{0.24\linewidth}
        \centering
        \includegraphics[width=\linewidth]{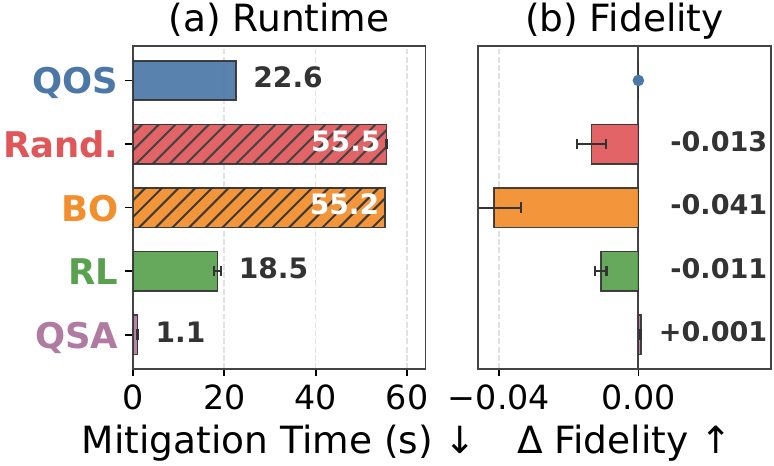}
    \caption{Impact of optimization strategies.}
    \label{fig:mitigation_ablation}
    \end{minipage}
\end{figure*}

\vspace{-10pt}
\subsection{Performance Breakdown}
\label{sec:performance_breakdown}

This subsection explains where \name's performance gains come from.
Additional evolution traces and case studies are provided in
Appendix~\cref{appendix:evolution_analysis} and
Appendix~\cref{appendix:case_study}.

\vspace{0.02in}
\noindent\textbf{Error-Mitigation Breakdown.}
Figure~\ref{fig:qsa_time_breakdown} decomposes end-to-end execution into
error mitigation, networking/scheduling, and per-circuit QPU computation.
As the circuit size increases from 12 to 24 qubits, mitigation time grows only
from 1.10\,s to 3.07\,s, while QPU computation increases by 21.3\%.
The key gain comes from the evolved mitigation logic: \name bounds the
fragment-size search, applies binary search and caching, and avoids unnecessary
expensive cutting probes, substantially reducing preprocessing overhead
(Appendix~\cref{appendix:case_study}).
Detailed depth and CNOT results are provided in
Appendix~\cref{appendix:circuit_property}.

\vspace{0.02in}
\noindent\textbf{Multiprogramming Policy Analysis.}
Figure~\ref{fig:figure_qsa_qos_mp_pareto_bar} explains the multiprogramming
gain through the quality of the evolved scheduling policy.
\name achieves an average Pareto rank of 5.59, substantially lower than QOS
and MP, and a policy correlation of 0.60, compared with 0.17 for QOS and
$-0.03$ for MP.
Thus, \name's gain comes from learning a policy that selects bundles closer to
the Pareto frontier while better preserving the ground-truth ordering, leading
to the improved fidelity--utilization trade-off observed in
Figure~\ref{fig:multiprogramming_result}.

\subsection{Ablation Study}
\label{sec:ablation_study}
This subsection quantifies the contribution of \name's harness components and examines its key design choices. Additional ablation studies can be found at Appendix~\ref{sec:additional_ablation}.

\vspace{0.02in}
\noindent\textbf{Impact of Approximation.}
We evaluate various approximation of real fidelity in multiprogramming evolution using the relative-balance ratio (\cref{sec:approx_scoring}) of: instruction count (Instr.), depth (Depth), non-local gates (NL), measurements (Measure), and CNOTs (see Appendix~\ref{appendix:approximation_detail} for details).
We also include simulatted fidelity (Sim. Fid.) for reference.
Figure~\ref{fig:impact_of_proxy} reveals that Instr., Depth, and NL collectively form the Pareto front of the utilization-fidelity trade-off.
While Instr. maximizes fidelity, Depth offers a superior balance, achieving $>96\%$ utilization with minimal fidelity loss.
At comparable utilization, Depth outperforms NL and CNOT as it more accurately reflects decoherence time, the dominant error source in NISQ devices.
Despite its higher cost, simulated fidelity does not improve the final fidelity--utilization trade-off.

\vspace{0.02in}
\noindent\textbf{Impact of Prompt Composition.}
Figure~\ref{fig:art_ablation} compares different prompt compositions in multiprogramming evolution.
We consider three components: 1) the Pareto ranks of the currently selected bundles (\cref{sec:snapshot_analysis}), 2) the system metrics of the ground-truth bundles (\cref{sec:snapshot_analysis}), and 3) the circuit properties of the ground-truth bundles (\cref{sec:static_analysis}).
We use a three-digit code to indicate whether each component is included.
Without static or snapshot analysis (000), both fidelity and normalized utilization are lower than the full design (111).
Partial prompt (001, 010, 100) improves over 000 and can approach 111, but is less stable.
Overall, the full composition (111) achieves the most stable and strongest performance, delivering high fidelity, maximum utilization, and minimal variance.

\vspace{0.02in}
\noindent\textbf{Impact of Policy Weight.}
In Figure~\ref{fig:score_ablation}, we assess the sensitivity of the system to the parameter $\alpha$ in the policy reward (Equation~\ref{eq:policy_reward}).
Keeping other settings fixed, we ablate $\alpha \in \{0, 0.2, 0.5, 0.8, 1.0\}$ in multiprogramming evolution.
When $\alpha=1$, \ie policy correlation is disabled, performance degrades noticeably in both fidelity and utilization.
When $\alpha=0$, \ie Pareto rank is disabled, fidelity also drops slightly.
Overall, the best performance is achieved by balancing Pareto rank and policy correlation, with $\alpha=0.5$ yielding the best evolution outcome.

\vspace{0.02in}
\noindent\textbf{Impact of Optimization Strategies.}
Figure~\ref{fig:mitigation_ablation} compares \name with Random, Bayesian
Optimization (BO), and Reinforcement Learning (RL) for error-mitigation design.
%
All methods target the same output decisions (target fragment size and cutting method
(GV or WC)) under a 60\,s search deadline.
These decisions follow the optimization interface of the QOS workflow.
%
Unlike conventional baselines that operate within predefined search or policy classes,
\name evolves the decision program itself.
Appendix~\ref{app:optimization_baselines}
details their configurations.
We report mitigation time and fidelity change relative to QOS, averaged across
circuit types and scales (12--24 qubits).
Random and BO frequently hit the deadline and are slower than QOS, while RL
provides only a modest speedup with a fidelity reduction of $0.011$.
%
%
The broader program-level search enables \name to discover decision strategies
beyond the predefined search or policy classes of conventional baselines.
Thus, \name reduces average mitigation time from 22.58\,s to
1.06\,s ($21.3\times$) while preserving fidelity, achieving the best overall
trade-off.

\begin{figure}[t]
    \centering
    \includegraphics[width=0.6\linewidth]{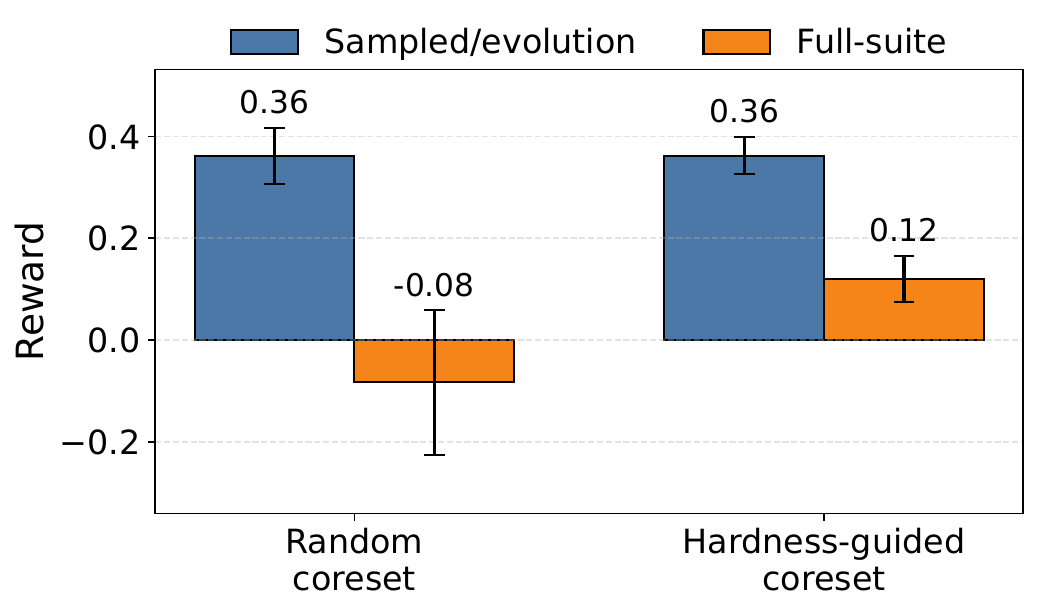}
    \caption{Impact of coreset selection on evolution reward.}
    \label{fig:randq_selection}
\end{figure}

\noindent\textbf{Impact of Coreset Selection.}
Figure~\ref{fig:randq_selection} compares uniform random coreset selection with our evolution-hardness-guided coreset under the same evolution setting and verification budget.
Both approaches achieve the same reward on the sampled verification set (0.36), but their full-suite performance differs substantially.
The random coreset yields a full-suite reward of $-0.08$, whereas the hardness-guided coreset improves it to $0.12$.
This result shows that uniform sampling can provide a misleading optimization signal, while focusing verification on an evolutionarily challenging circuit scale produces a more effective coreset for guiding evolution.

\section{Discussion}
This section discusses the automation scope of \name and its generalizability.
More discussions are at Appendix~\cref{appendix:additional_discussion}.

\vspace{0.02in}
\noindent\textbf{Automation Scope.}
\name is semi-automated: engineers provide a one-time task specification,
including the target software component, an initial implementation, and the
optimization objectives.
Given this specification, the quantum-aware harness automatically drives the
subsequent evolution process by generating mutations, verifying candidate
programs, collecting quantum-specific execution context, and providing rewards
to guide further mutations.
Full automation could further let the LLM formulate the initial implementation
and optimization specification, but reliably exploring such an unconstrained
software design space remains challenging for current LLMs~\cite{ma2023eureka,xia2024agentless}.
\name therefore constrains the search to a specified software component while
automating its iterative optimization.

\vspace{0.02in}
\noindent\textbf{Hardware Generalizability.}
\name is not tied to a specific quantum backend (\eg the IBM Quantum platform).
Its harness relies on observable circuit properties, execution states, and
system metrics rather than assumptions about a particular hardware
architecture.
Therefore, \name can be applied to different or future quantum platforms as
long as the metrics required by the optimization objective can be measured and
the execution stack exposes sufficient feedback for verification and prompting.
As quantum hardware exposes more software-configurable controls
(\eg circuit layering or pulse-level parameters), the same evolutionary
approach could further extend to lower-level system optimizations.

\section{Related Work}
\noindent\textbf{Error Mitigation.}
Many techniques mitigate NISQ noise across different abstraction layers.
Zero-Noise Extrapolation estimates noise-free expectations by scaling noise and extrapolating to the zero-noise limit~\cite{li2017efficient,temme2017error}, while Probabilistic Error Cancellation reconstructs ideal outcomes through quasi-probability sampling~\cite{endo2018practical,temme2017error}.
At the pulse and circuit levels, Dynamical Decoupling~\cite{das2021adapt,pokharel2018demonstration} and Pauli Twirling~\cite{wallman2016noise} reduce decoherence and coherent noise.
Readout Error Mitigation corrects measurement noise through hardware calibration~\cite{bravyi2021mitigating}.
At the system level, recent frameworks such as QOS incorporate noise-aware mapping and scheduling based on hardware calibration.
\name automates the discovery of error mitigation strategies through compiler-pass evolution, uncovering better fidelity--latency trade-offs.

\vspace{0.02in}
\noindent\textbf{Multiprogramming.}
Quantum multiprogramming evolved from rigid physical isolation to suppress crosstalk~\cite{das2019case} toward flexible mapping techniques like X-SWAP~\cite{liu2021qucloud} that improve spatial utilization by relaxing program boundaries.
QOS~\cite{giortamis2025qos} uses admission control for fidelity-resource balancing, while QVM~\cite{tao2025quantum} leverages virtualization to enforce execution isolation.
However, these systems rely on static heuristics that struggle to navigate complex trade-offs between fidelity and utilization as hardware scales increase.
\name bridges this gap by abstracting decision logic into evolvable policies, autonomously synthesizing scheduling rules that approach the Pareto-optimal frontier.

\vspace{0.02in}
\noindent\textbf{LLM-Based System Optimization.}
Traditional approaches optimize systems through Bayesian optimization~\cite{snoek2012practical,frazier2018tutorial}, RL-based fine-tuning~\cite{mirhoseini2017device,zhou2020transferable}, and program synthesis~\cite{gulwani2017program,solar2008program}.
However, these approaches typically require predefined parameter spaces, action spaces, or synthesis constraints, making them less suitable for optimizing complex structural code~\cite{cummins2024meta}.
Recent work therefore increasingly leverages LLMs to directly generate and revise system code~\cite{lin2025byos,park2025principles,cummins2024meta}.
LLM-guided evolutionary search further enables iterative refinement by repeatedly generating, evaluating, and improving candidate programs.
Evolutionary discovery of bivariate bicycle codes~\cite{cruzbenito2026evolutionary} applies this paradigm to quantum code discovery, but focuses on a specific quantum artifact.
In contrast, \name targets reusable quantum software components, including compiler passes and runtime policies.
More importantly, existing evolutionary search frameworks provide only a generic search loop and lack the quantum-specific mechanisms required for efficient verification, informative execution feedback, and heterogeneous reward design.
\name addresses this gap with a \emph{quantum-aware harness} that equips generic LLM-guided evolutionary search with these domain-specific mechanisms.

\section{Conclusion}
Rapid advances in quantum hardware call for automated software design to keep pace with evolving systems and move beyond handcrafted heuristics.
We present \name, a \emph{quantum-aware harness} that enables LLM-guided evolutionary search to automatically optimize quantum compiler passes and runtime policies.
By providing quantum-specific mechanisms for efficient verification, informative execution feedback, and task-specific reward design, \name makes generic evolutionary search practical for quantum software.
Our evaluation shows that \name consistently outperforms SOTA baselines across fidelity, QPU utilization, and execution time, demonstrating the potential of quantum-aware evolutionary search for automated quantum software engineering.

\newpage

\bibliographystyle{ACM-Reference-Format}
\bibliography{reference}

\appendix
\section{IBM Quantum Platform Analysis}
\label{sec:ibm_logging}

This section provides additional characterization of resource availability and
hardware noise on the IBM Quantum platform, complementing the discussion in
\S\ref{sec:quantum_software_design}.

\begin{figure}[t]
  \begin{minipage}[t]{0.46\linewidth}
    \centering
    \includegraphics[width=\linewidth]{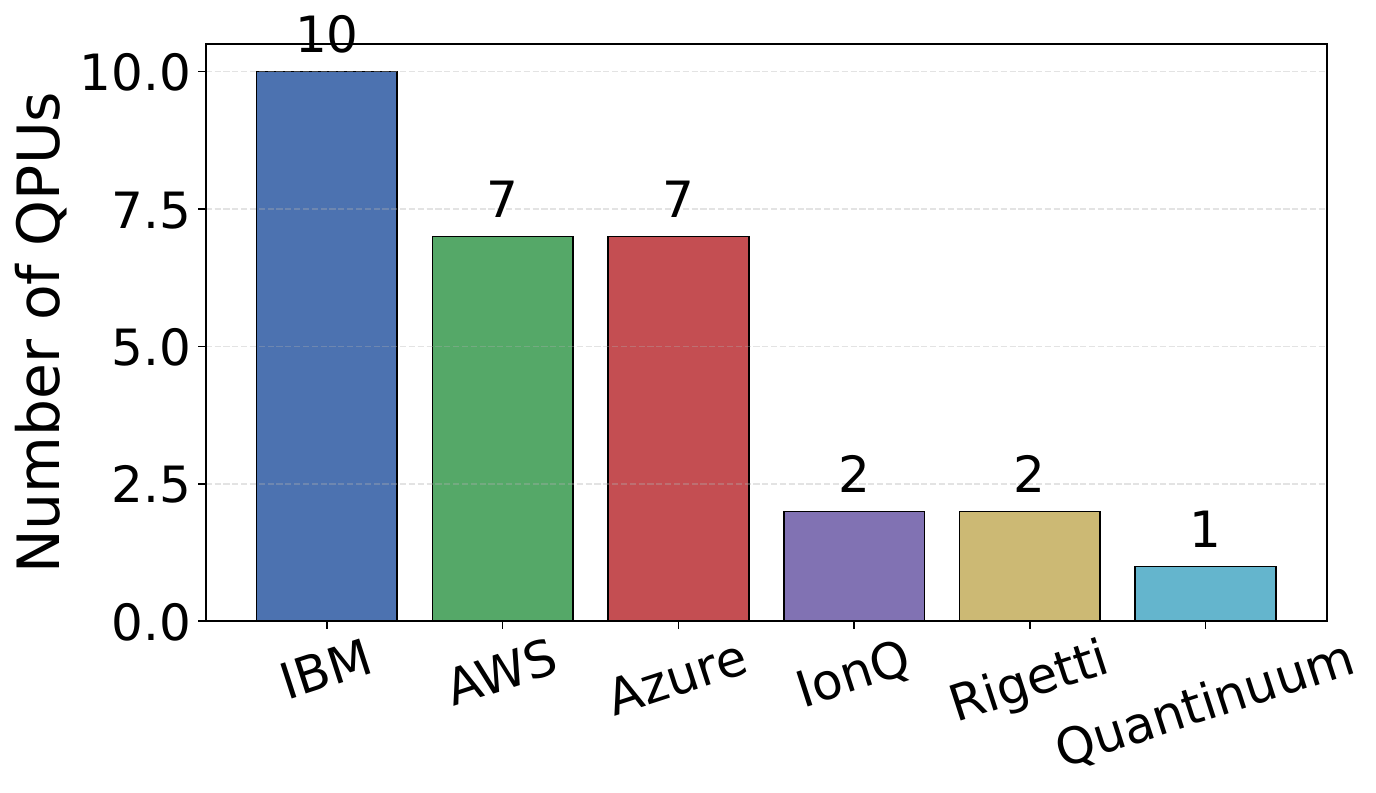}
    \caption{Publicly accessible QPUs across major cloud providers.}
    \label{fig:cloud_accessible_qpus}
  \end{minipage}
  \hfill
  \begin{minipage}[t]{0.46\linewidth}
    \centering
    \includegraphics[width=\linewidth]{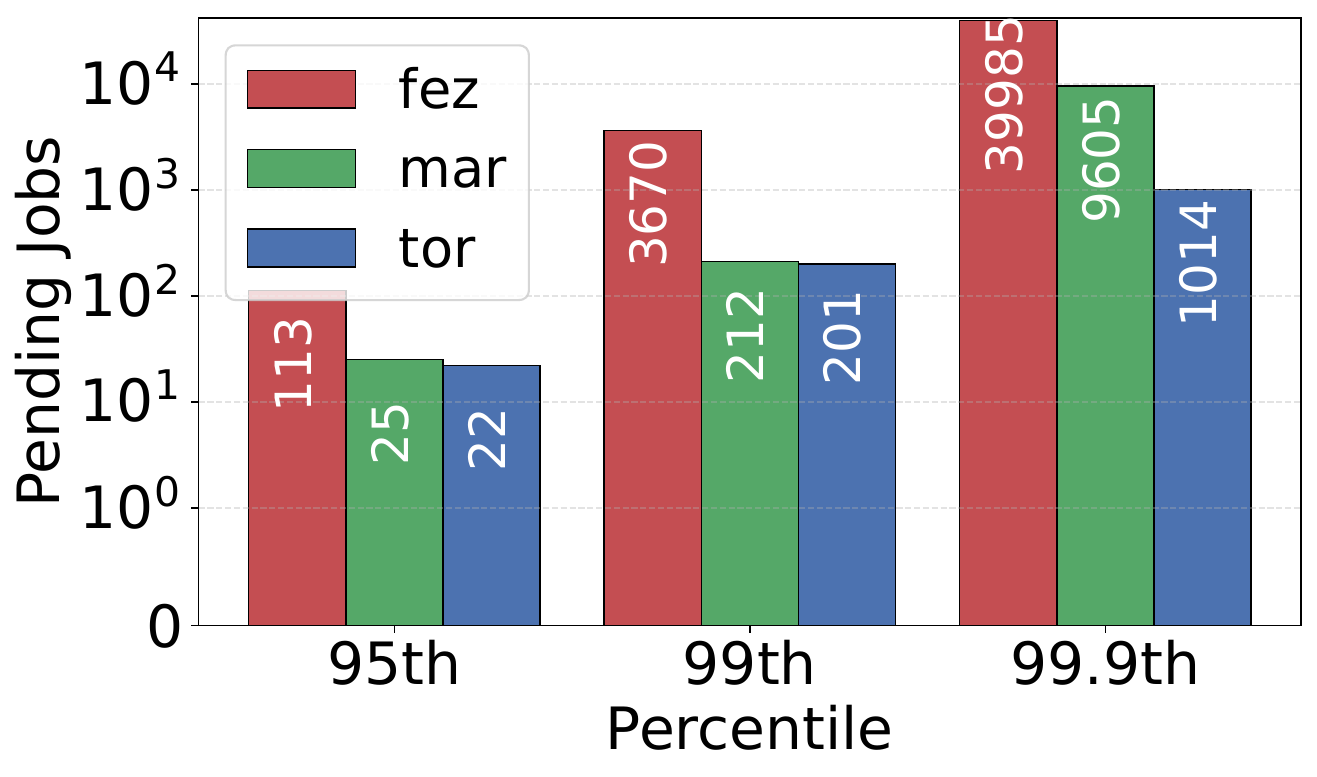}
    \caption{High-percentile pending-job counts on IBM QPUs.}
    \label{fig:pending_jobs}
  \end{minipage}
\end{figure}

\noindent\textbf{Resource Availability.}
Figure~\ref{fig:cloud_accessible_qpus} shows that public quantum clouds expose
only a limited number of QPUs.
To further characterize resource contention, we collect the number of pending
jobs on IBM\_torino, IBM\_marrakesh, and IBM\_fez every 10 minutes throughout
March 2026.
Figure~\ref{fig:pending_jobs} reports the 95th, 99th, and 99.9th percentiles of
the observed queue lengths.
The long-tail queueing behavior is substantial; for example, the
99.9th-percentile queue length of IBM\_fez reaches 39{,}985 jobs.

\begin{figure}[t]
    \centering
    \includegraphics[width=\linewidth]{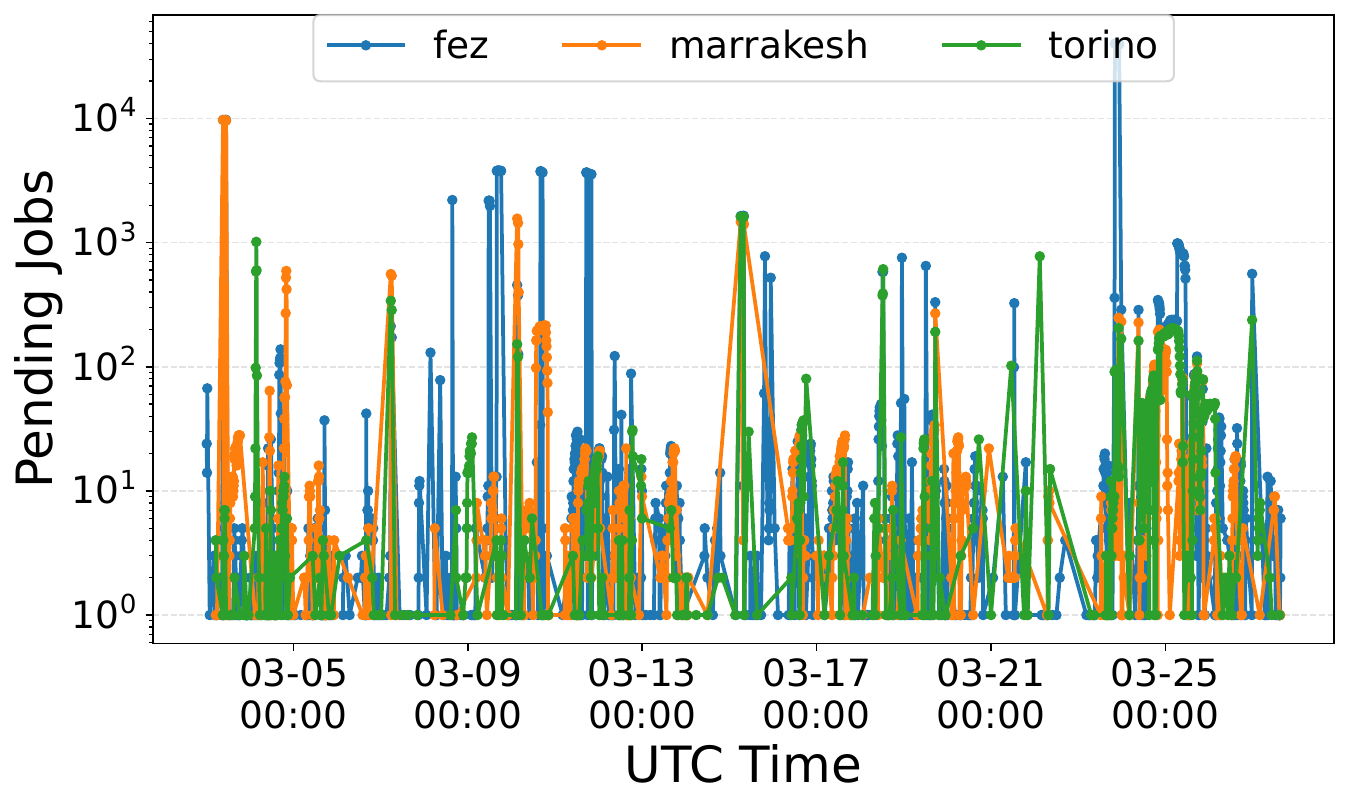}
    \caption{Pending jobs on IBM QPUs throughout March 2026.}
    \label{fig:pending_jobs_over_time_log}
\end{figure}

Figure~\ref{fig:pending_jobs_over_time_log} further shows how queue lengths
vary over time.
IBM\_fez consistently experiences heavier contention than the other observed
QPUs, while all three exhibit substantial temporal variation.

\begin{table}[t]
\centering
\begin{tabular}{lccc}
\toprule
Backend & 2Q Gate ($\mu$) & 1Q Gate ($\mu$) & Readout ($\mu$)\\
\midrule
\rowcolor{red!10}
Torino &
$2.564\times10^{-3}$ &
$3.172\times10^{-4}$ &
$\mathbf{2.961\times10^{-2}}$ \\
Marrakesh &
$2.632\times10^{-3}$ &
$3.201\times10^{-4}$ &
$1.160\times10^{-2}$ \\
\rowcolor{green!10}
Fez &
$2.626\times10^{-3}$ &
$2.846\times10^{-4}$ &
$1.462\times10^{-2}$ \\
\bottomrule
\end{tabular}
\caption{Representative error metrics of the evaluated IBM QPUs.}
\label{tab:ibm_qpu_error_stats}
\end{table}

\noindent\textbf{Hardware Noise.}
Table~\ref{tab:ibm_qpu_error_stats} summarizes representative 1-qubit,
2-qubit, and readout error rates for the three IBM QPUs.
The backends exhibit different error characteristics: IBM\_torino has the
highest readout error, while IBM\_fez has the lowest 1-qubit gate error among
the measured systems.

\begin{figure}[t]
    \centering
    \includegraphics[width=\linewidth]{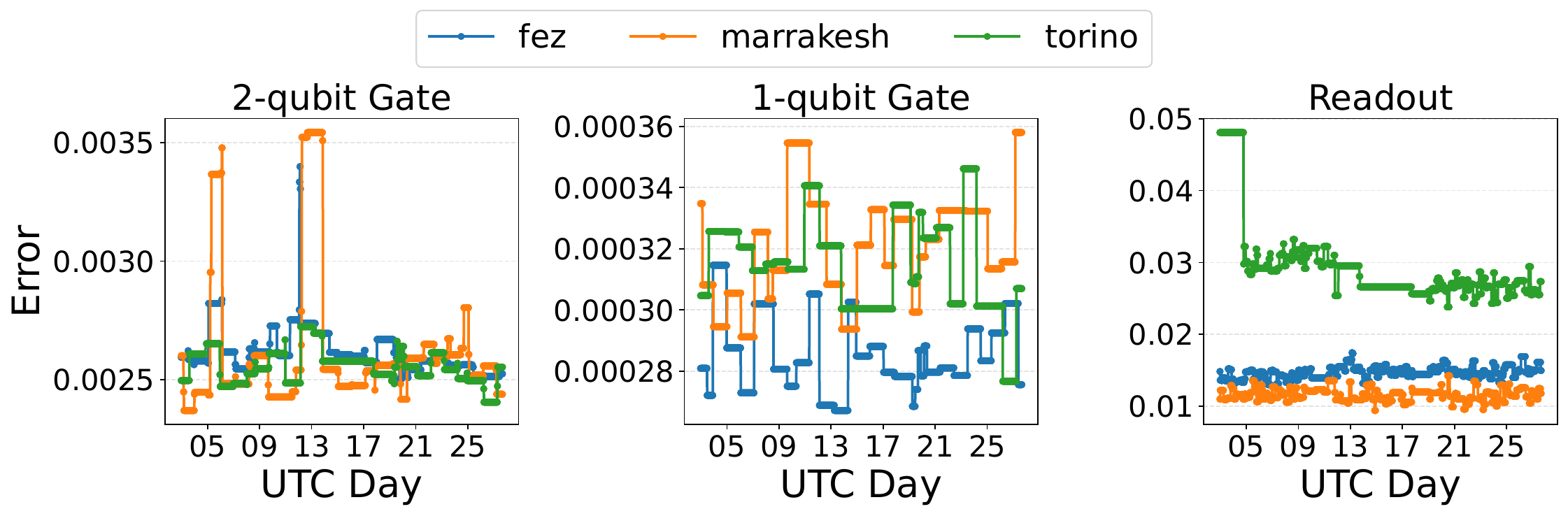}
    \caption{Variation of IBM QPU error metrics over time.}
    \label{fig:error_metrics_over_time}
\end{figure}

Figure~\ref{fig:error_metrics_over_time} shows that these error metrics also
vary over time.
The 2-qubit gate error rates are broadly comparable across the observed QPUs,
although IBM\_marrakesh exhibits greater temporal variation.
IBM\_fez generally has the lowest 1-qubit gate error, whereas IBM\_torino
exhibits the highest readout error.
Together, these measurements highlight both resource contention and
backend-dependent noise variation in today's quantum cloud environments.

\begin{table*}[t]
\centering
\small
\setlength{\tabcolsep}{5pt}
\begin{tabularx}{\linewidth}{l X}
\toprule
\textbf{Static Features} & \textbf{Physical Meaning} \\
\midrule
\texttt{depth} & Number of sequential gate layers in the circuit. \\
\texttt{num\_qubits} & Number of qubits used by the circuit. \\
\texttt{num\_clbits} & Number of classical bits used by the circuit. \\
\texttt{num\_nonlocal\_gates} & Number of nonlocal gates, \ie a two-qubit gate between qubits that are not directly connected on the hardware coupling graph, so SWAPs~\cite{wille2014optimal} must first move their states together. \\
\texttt{num\_connected\_components} & Number of disconnected components in the qubit-interaction graph. \\
\texttt{number\_instructions} & Total number of instructions in the circuit. \\
\texttt{num\_measurements} & Number of measurement operations. \\
\texttt{num\_cnot\_gates} & Number of CNOT (controlled-X) gates. \\
\texttt{program\_communication} & Degree of interaction or dependency among different parts of the circuit. \\
\texttt{liveness} & Degree to which qubits remain active throughout circuit execution. \\
\texttt{parallelism} & Degree to which operations can be executed concurrently. \\
\texttt{measurement} & Relative amount or intensity of measurement operations in the circuit. \\
\texttt{entanglement\_ratio} & Fraction of entangling gates among all gates. \\
\texttt{critical\_depth} & Length of the longest dependency chain in the circuit. \\
\bottomrule
\end{tabularx}
\caption{Physical meaning of static features used in the evolution target.}
\label{tab:static_features}
\end{table*}

\section{Static Feature Details}
\label{appendix:static_feature_details}
Table~\ref{tab:static_features} describes the 14 static features extracted through static analysis and their physical meanings.
It combines structural size/complexity features (\eg depth, qubit count, instruction/CNOT/nonlocal-gate counts), graph/connectivity features (\eg connected components), and execution-shape features (\eg parallelism, liveness, measurement intensity, entanglement ratio, critical depth).
These features give a compact physical profile of each circuit, so the optimizer can prioritize candidates that are likely easier to mitigate while preserving fidelity and runtime efficiency.

\begin{table}[t]
\centering
\small
\begin{tabularx}{\linewidth}{p{0.30\linewidth} X}
\toprule
\textbf{Block} & \textbf{Content} \\
\midrule
Format Req & Editable region and function signature. \\
Verifier Spec & Scoring formula description. \\
Parent Program & Parent implementation to mutate. \\
History Program & Top/Diverse/Inspirational programs. \\
+ Static Analysis & Extracted features through analysis. \\
+ Snapshot Analysis & High-level scores and per-case traces. \\
\bottomrule
\end{tabularx}
\caption{Prompt description.}
\label{tab:prompt_block_summary}
\end{table}
\section{OpenEvolve Configurations}
\label{appendix:configurations}
%
Listings~\ref{lst:oe_config_em} and \ref{lst:oe_config_mp} present the OpenEvolve configurations for error mitigation and multiprogramming, respectively.
In addition to the static and snapshot analysis of our design, they include four basic components from OpenEvolve~\cite{openevolve}: 
(1) \emph{Format requirement}, which specifies editable code regions and function signatures to ensure syntactic correctness; 
(2) \emph{Verifier specification}, which defines the reward formulation (\eg a single metric or a weighted sum of multiple metrics); 
(3) \emph{Parent program}, selected via a stochastic policy from top-quality candidates (exploitation) or random candidates (exploration); and 
(4) \emph{Historical programs}, where top programs guide optimization toward high-quality solutions, diverse programs broaden exploration, and inspiration programs provide reusable patterns for generation. 
We summarize them in Table~\ref{tab:prompt_block_summary}.

\begin{figure}
    \centering
    \includegraphics[width=\linewidth]{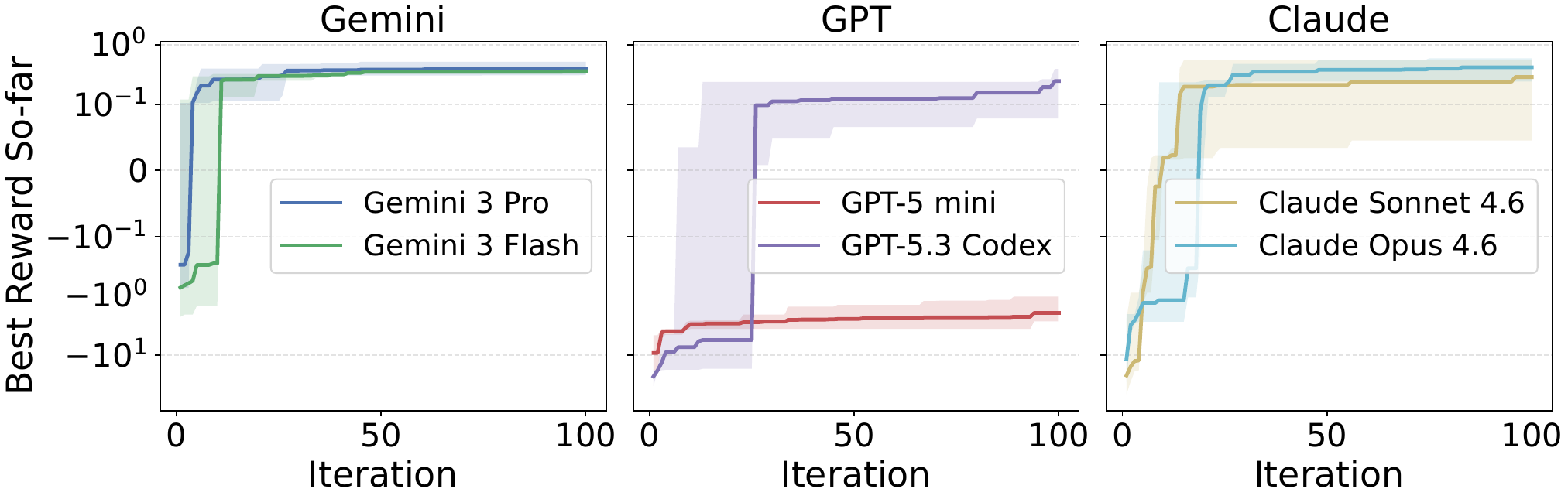}
    \caption{Learning curves for error mitigation.}
    \label{fig:learning_curves_em}
\end{figure}

\section{Evolution Analysis}
\label{appendix:evolution_analysis}
\begin{figure}[t]
    \centering
    \includegraphics[width=\linewidth]{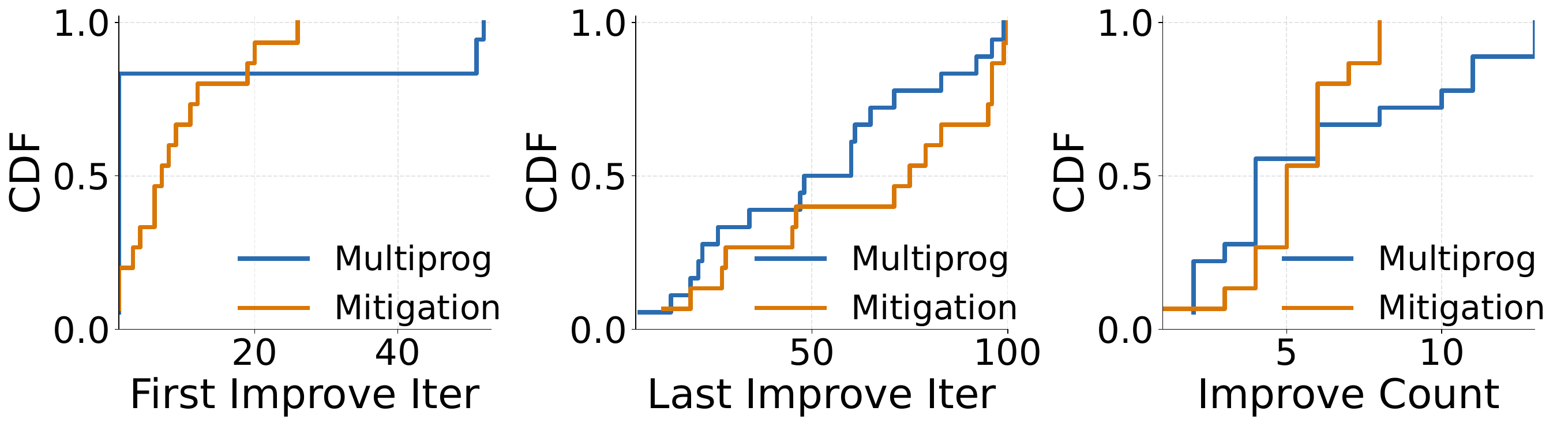}
    \caption{Improvement analysis.}
    \label{fig:improvement_cdfs_all_models}
\end{figure}
\noindent\textbf{Overall.}
We analyze all evolution runs that achieve at least one improvement within 100 iterations for both multiprogramming and error mitigation.
Figure~\ref{fig:improvement_cdfs_all_models} shows the CDFs of the first improvement iteration, the last improvement iteration, and the total number of improvements.
The first improvement typically occurs within 27 iterations, suggesting that improvement tends to happen early or not at all.
For multiprogramming, the first improvement appears within just 2 iterations, as it is an easier target to optimize.
In contrast, the last improvement may occur at any point within the 100 iterations, highlighting the stochastic nature of evolution.
The number of improvements is broadly distributed between 0--8 for multiprogramming and 0--15 for error mitigation, indicating sparse reward-improving mutations.
Detailed learning curves are explained in the following paragraph.

\noindent\textbf{Learning Curves.}
Figure~\ref{fig:learning_curves_em} illustrates the learning curve for Error Mitigation using different LLM APIs.
Gemini models are the first to improve the best reward.
Claude models are the second and GPT models are the last.
Notably, \GPTF can hardly improve the reward.
In terms of the reward value, Claude achieves the highest score.
Unfortunately, it fails to generalize well as described in Figure~\ref{fig:mitigation_model_comparison}.

\begin{figure}
    \centering
    \includegraphics[width=\linewidth]{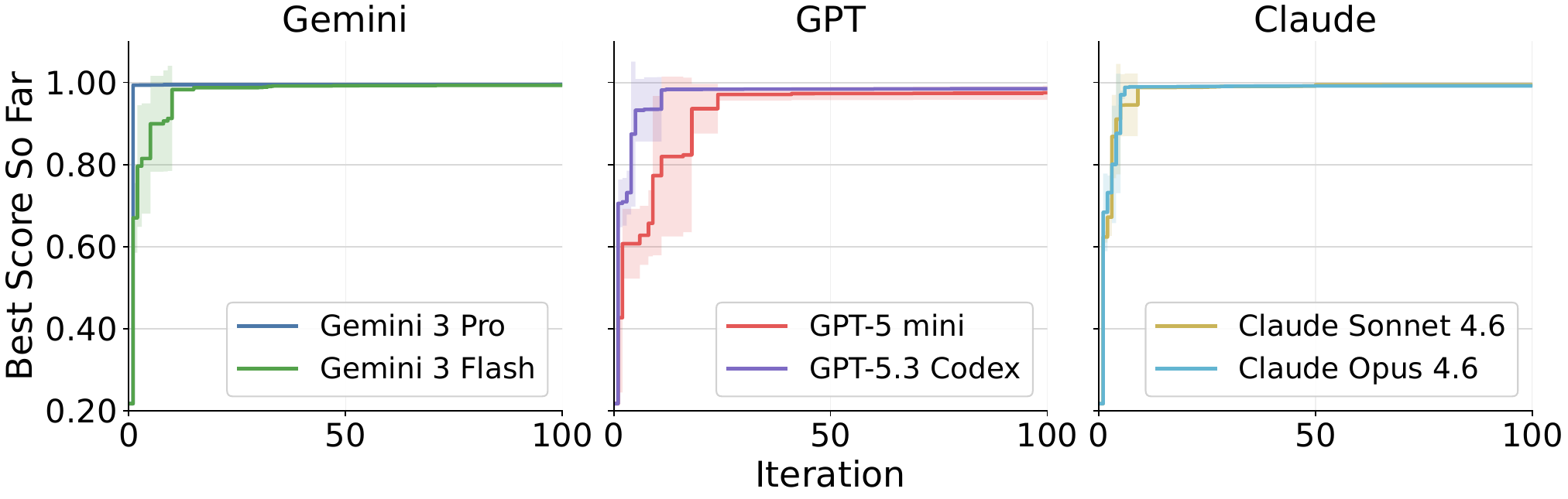}
    \caption{Learning curves for multiprogramming.}
    \label{fig:learning_curves_mp}
\end{figure}

Figure~\ref{fig:learning_curves_mp} shows the learning curve for Multiprogramming using different LLM APIs.
Most of the reward gain is obtained in the early iterations, while later iterations mainly provide marginal refinements.
Gemini 3 Pro converges the fastest, reaching near-final performance almost immediately.
In addition, Multiprogramming exhibits earlier and more frequent improvements than Error Mitigation.
It suggesting that the search space in the former setting is easier for LLM-driven optimization.

\section{Case Study}
\label{appendix:case_study}
\subsection{Best Case Analysis.}
Our evolution using \GeminiTP results in a sophisticated design for error mitigation but a simple-yet-effective one for multiprogramming.
For error mitigation, \name derives a bounded cost-search policy: it does binary search over target size, caches probe results, prefers gate virtualization first, and only pays for wire cutting checks when likely useful. This approach gives a strong runtime cut while preserving circuit properties (depth/CNOT).
For multiprogramming, \name distills a minimalist heuristic by discarding 14+ features in favor of a two-variable model: utilization and depth ratio. 
This selection minimizes idle-time decoherence while ensuring near-zero scheduling overhead.
More details can be found at Appendix~\ref{appendix:initial_best_programs}.

\begin{figure}
    \centering
    \includegraphics[width=0.6\linewidth]{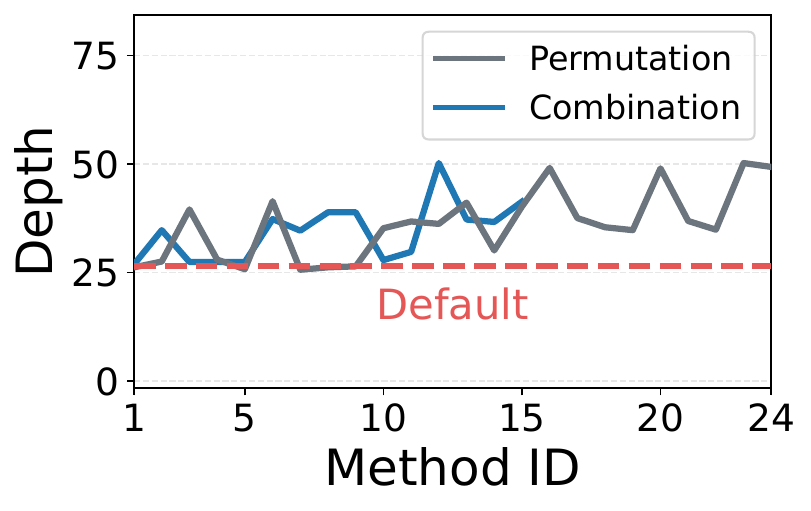}
    \caption{Analysis of combination and permutation.}
    \label{fig:methods_combo_perm_all_12_aggregate_depth}
\end{figure}

\subsection{Failed Case Analysis.}
Effective abstraction is critical for successful policy evolution.
%
%
We initially had an hypothesis that a different combination or permutation of techniques~\cite{ayanzadeh2023frozenqubits,mitarai2021constructing,peng2020simulating,decross2023qubit,jiang2024qubit} adopted by QOS~\cite{giortamis2025qos} may yield a better performance.
%
We then abstract the details of these techniques (including circuit cutting that should have exposed the cost search logic) and tasked the LLM with exploring various permutations and combinations.
This approach yielded no measurable improvements because operator ordering has negligible impact on final mitigation efficacy.
As analyzed in Figure~\ref{fig:methods_combo_perm_all_12_aggregate_depth}, neither combination nor permutation outperform the default setting.
Overall, successful evolution requires precise abstraction driven by solid weakness analysis results.

\subsection{Initial and Best Programs}
\label{appendix:initial_best_programs}
Listing~\ref{lst:initial_program_em} and Listing~\ref{lst:best_program_em} show the initial program and the best program (generated by Gemini 3 Pro) for the error mitigation task.
The best program performs well because it turns cost-search into a bounded, efficient decision process: it uses metadata to set a valid search range, applies a binary search to find the smallest feasible target size under budget, and avoids unnecessary expensive probes (especially wire cutting) during the search. That usually yields smaller feasible fragment sizes (helping depth/CNOT proxies) while reducing search overhead. It also includes caching and a final verification step, which improves stability and keeps runtime predictable across circuits.
%
Listing~\ref{lst:initial_program_mp} and Listing~\ref{lst:best_program_mp} show the initial program and the best program (generated by Gemini 3 Pro) for the error mitigation task.
The evolved program identifies a minimalist, principle-driven heuristic. 
It condenses 14+ architectural features into a weighted sum of effective utilization ($0.65$) and depth similarity ($0.35$).
This shift validates a key system insight: quantum resource management is fundamentally a temporal synchronization problem.
By prioritizing the depth ratio, the logic enforces temporal alignment. 
%
%
Such policy parsimony ensures near-zero scheduling overhead and robust generalization across diverse hardware backends.

\begin{lstlisting}[
  basicstyle=\footnotesize\ttfamily,
  breaklines=true,
  breakatwhitespace=false,
  columns=fullflexible,
  frame=single,
  caption={OpenEvolve Configuration for Error Mitigation},
  label={lst:oe_config_em},
  captionpos=t,
]
max_iterations: 100
checkpoint_interval: 10
log_level: "INFO"
log_dir: null
random_seed: 42
max_tasks_per_child: 1

diff_based_evolution: false
max_code_length: 30000

early_stopping_patience: null
convergence_threshold: 0.001
early_stopping_metric: "combined_score"

llm:
  api_base: "${OPENAI_API_BASE}"
  api_key: "${OPENAI_API_KEY}"
  primary_model: "${OPENAI_MODEL}"
  temperature: 0.7
  top_p: 0.9
  max_tokens: 4096
  timeout: 600
  retries: 3
  retry_delay: 15

prompt:
  system_message: |
    You are optimizing ONLY the function evolved_cost_search(self, q, size_to_reach, budget)
    in qos/error_mitigator/evolution_target.py.

    OUTPUT FORMAT (CRITICAL):
    - Return the full file contents of qos/error_mitigator/evolution_target.py.
    - Output code only: no explanations, no markdown fences, no extra text.

    GENERALIZATION REQUIREMENT (CRITICAL):
    - The evaluator samples random (circuit_type, num_qubits) pairs each run.
    - Qubit counts are constrained to [12,24] and must exist in the benchmark set.
    - Your logic must generalize across circuit families (QAOA, BV, GHZ, VQE, QSVm, etc.) and sizes.
    - Do NOT branch on q.get_name(), bench strings, or any per-benchmark constants.

    FUNCTION CONTRACT (MUST FOLLOW EXACTLY):
    - Do NOT change the function signature.
      It MUST remain: evolved_cost_search(self, q, size_to_reach, budget)
    - It MUST return EXACTLY 2 values: (size_to_reach: int, method: str) where method is "GV" or "WC".
    - If you violate signature/return arity, evaluation will fail.

    CONTEXT:
    The input q is a Qernel. It contains metadata computed elsewhere, accessible via:
      input_meta = q.get_metadata()

    Useful fields may include:
      depth, num_qubits, num_clbits, num_nonlocal_gates, num_connected_components,
      number_instructions, num_measurements, num_cnot_gates,
      program_communication, liveness, parallelism, measurement,
      entanglement_ratio, critical_depth.

    COST HELPER APIS (from qos/error_mitigator/run.py):
      - compute_gv_cost(q, size_to_reach, timeout_sec=0)
      - compute_wc_cost(q, size_to_reach, timeout_sec=0)
      - Each call returns (cost: int, timed_out: bool).
      - timeout_sec is per-call; GV/WC can use different values.
      - timeout_sec=0 means no per-call timeout.
      - On timeout, cost is set high and timed_out=True.

    Inputs/Outputs:
      - Inputs:
          * q: the Qernel (quantum circuit container) whose metadata informs the decision.
          * size_to_reach: the current target circuit size for cutting/virtualization search.
          * budget: the cost threshold used to decide whether GV/WC is acceptable.
      - Outputs:
          * size_to_reach: the chosen target size to use for the cutting/virtualization of the target circuit.
          * method: "GV" or "WC", which determines which technique is applied at that size.
      - Cost logging and traces are handled outside this file (in qos/error_mitigator/run.py).

    IMPORTANT PRACTICAL CONSTRAINTS:
    - `budget` is only meaningful for your cost-search logic (i.e., your own GV/WC cost queries).
      Do not assume downstream will enforce budget.
    - If size_to_reach >= num_qubits, the decomposition can become a no-op.
      Use that ONLY as an explicit fallback when costs indicate everything else is too expensive.
      Prefer to clamp sizes into a meaningful range (e.g., [2, num_qubits-1]) for real action.

    EFFICIENCY / GENERALIZATION GUIDELINES (HIGH IMPACT):
    - Minimize gv/wc cost calls. Avoid long +/-1 stepping loops.
      Prefer a small number of probes + bounded search (e.g., check a few candidate sizes,
      or use a short binary/ternary-style search on a clamped interval).
    - Use cheap metadata heuristics first; only call compute_*_cost when needed.
    - Keep behavior monotone and stable: similar circuits should yield similar decisions.

    NOTE: "QOS" refers to the baseline implementation, and "QOSE" refers to the evolved program.

    METRICS:
      - qose_depth: average depth ratio (QOSE/QOS)
      - qose_cnot: average CNOT gate ratio (QOSE/QOS)
      - qose_overhead: average circuit-count ratio (QOSE/QOS)
      - avg_run_time: average runtime ratio (QOSE/QOS)
      - combined_score (maximize): evaluator score used by evolution ranking.
        combined_score = - (qose_depth + qose_cnot + 1./4 * avg_run_time)
    NOTE:
      - Depth/CNOT ratios are proxies for fidelity. The goal is to improve fidelity,
        reduce cost-search time, and avoid depth/CNOT regressions.

    We provide the execution outputs of the last program executed by the evaluator. 
    Use them to diagnose decisions. Details of the execution outputs are as follows:

    EXECUTION OUTPUTS PROVIDED BY EVALUATOR:
      - Artifacts are organized as:
        * artifacts["summary"] (optional): aggregate/run-level metrics
        * artifacts["cases"] (optional): list of per-(bench,size) case metrics

    EXECUTION OUTPUTS PROVIDED BY EVALUATOR (artifacts["summary"]):
      - qose_budget: budget used for all cases
      - qose_run_sec_avg: average runtime of evolved mitigator
      - qos_run_sec_avg: average runtime of baseline QOS
      - gv_cost_calls_total: total GV cost calls (evolved)
      - wc_cost_calls_total: total WC cost calls (evolved)
      - qos_gv_cost_calls_total: total GV cost calls (baseline)
      - qos_wc_cost_calls_total: total WC cost calls (baseline)
      - qose_depth_avg: average absolute depth of evolved circuits
      - qose_cnot_avg: average absolute CNOT gate count of evolved circuits
      - qose_overhead_avg: average absolute number of circuits generated by evolved method
      - qos_depth_avg: average absolute depth of baseline circuits
      - qos_cnot_avg: average absolute CNOT gate count of baseline circuits
      - qos_overhead_avg: average absolute number of circuits generated by baseline

    EXECUTION OUTPUTS PROVIDED BY EVALUATOR (artifacts["cases"] per case):
      - input_features: static circuit features (depth, qubits, gates, etc.)
        num_connected_components, number_instructions, num_measurements, num_cnot_gates,
        program_communication, liveness, parallelism, measurement, entanglement_ratio, critical_depth
      - qose_input_size: initial target size before cost-search adjustments
      - qose_depth, qos_depth, qose_cnot, qos_cnot,
        qose_run_sec, qos_run_sec: paired QOSE vs baseline QOS metrics
      - qose_output_size: final target size chosen by the algorithm
      - qose_method: chosen method ("GV" or "WC")
      - qose_gv_cost_trace, qose_wc_cost_trace: cost estimates per probe step
      - qose_gv_time_trace, qose_wc_time_trace: time per probe step

    POSSIBLE LEVERS TO EXPLORE:
      A) Decide to use WC or GV selectively instead of always using both.
      B) Explore size_to_reach efficiently instead of using two while loops iteratively.
      C) Explore early stopping rules for WC and GV.
      D) Skip certain size_to_reach values based on heuristics.
      

  evaluator_system_message: "You are a strict code reviewer. Do not add commentary."
  num_top_programs: 3
  num_diverse_programs: 2
  use_template_stochasticity: true

  # Keep artifact injection enabled by default.
  include_artifacts: true
  max_artifact_bytes: 65536
  artifact_security_filter: true

database:
  db_path: null
  in_memory: true
  log_prompts: true

  population_size: 100
  archive_size: 20
  num_islands: 3

  migration_interval: 30
  migration_rate: 0.12

  elite_selection_ratio: 0.05
  exploration_ratio: 0.65
  exploitation_ratio: 0.30

  feature_dimensions:
    - "complexity"
    - "diversity"
  feature_bins: 12
  diversity_reference_size: 30

evaluator:
  timeout: 12000
  max_retries: 1
  cascade_evaluation: false
  parallel_evaluations: 1
  use_llm_feedback: false
  llm_feedback_weight: 0.0

evolution_trace:
  enabled: false
  format: "jsonl"
  include_code: false
  include_prompts: false
  output_path: null
  buffer_size: 10
  compress: false

\end{lstlisting}

\begin{lstlisting}[
  language=Python,
  basicstyle=\footnotesize\ttfamily,
  numbers=left,
  breaklines=true,
  breakatwhitespace=false,
  columns=fullflexible,
  frame=single,
  caption={Initial Program for Error Mitigation.},
  label={lst:initial_program_em},
  captionpos=t,
]
from qos.error_mitigator.run import compute_gv_cost, compute_wc_cost
from qos.types.types import Qernel


def evolved_cost_search(self, q: Qernel, size_to_reach: int, budget: int):
    metadata = q.get_metadata()
    depth = metadata.get("depth", 0)
    num_qubits = metadata.get("num_qubits", 0)
    num_clbits = metadata.get("num_clbits", 0)
    num_nonlocal_gates = metadata.get("num_nonlocal_gates", 0)
    num_connected_components = metadata.get("num_connected_components", 0)
    number_instructions = metadata.get("number_instructions", 0)
    num_measurements = metadata.get("num_measurements", 0)
    num_cnot_gates = metadata.get("num_cnot_gates", 0)
    program_communication = metadata.get("program_communication", 0.0)
    liveness = metadata.get("liveness", 0.0)
    parallelism = metadata.get("parallelism", 0.0)
    measurement = metadata.get("measurement", 0.0)
    entanglement_ratio = metadata.get("entanglement_ratio", 0.0)
    critical_depth = metadata.get("critical_depth", 0.0)

    # OE_BEGIN
    gv_timeout_sec = 0
    wc_timeout_sec = 0
    gv_cost, _gv_timed_out = compute_gv_cost(q, size_to_reach, timeout_sec=gv_timeout_sec)
    wc_cost, _wc_timed_out = compute_wc_cost(q, size_to_reach, timeout_sec=wc_timeout_sec)
    while (gv_cost <= budget or wc_cost <= budget) and size_to_reach > 2:
        size_to_reach = size_to_reach - 1
        gv_cost, _gv_timed_out = compute_gv_cost(
            q, size_to_reach, timeout_sec=gv_timeout_sec
        )
        wc_cost, _wc_timed_out = compute_wc_cost(
            q, size_to_reach, timeout_sec=wc_timeout_sec
        )

    while gv_cost > budget and wc_cost > budget:
        size_to_reach = size_to_reach + 1
        gv_cost, _gv_timed_out = compute_gv_cost(
            q, size_to_reach, timeout_sec=gv_timeout_sec
        )
        wc_cost, _wc_timed_out = compute_wc_cost(
            q, size_to_reach, timeout_sec=wc_timeout_sec
        )

    method = "GV" if gv_cost <= wc_cost else "WC"
    # OE_END
    return size_to_reach, method
\end{lstlisting}

\begin{lstlisting}[
  language=Python,
  basicstyle=\footnotesize\ttfamily,
  numbers=left,
  breaklines=true,
  breakatwhitespace=false,
  columns=fullflexible,
  frame=single,
  caption={Best Program by Gemini-3-Pro for Error Mitigation},
  label={lst:best_program_em},
  captionpos=t,
]
from qos.error_mitigator.run import compute_gv_cost, compute_wc_cost
from qos.types.types import Qernel

def evolved_cost_search(self, q: Qernel, size_to_reach: int, budget: int):
    # OE_BEGIN
    # Optimization Strategy:
    # 1. Goal: Minimize 's' (subcircuit size) such that Cost(s) <= budget.
    #    Smaller 's' -> more cuts -> better fidelity proxy (lower depth/CNOTs).
    # 2. Search: Binary search on [2, num_qubits].
    # 3. Efficiency:
    #    - Prioritize GV (usually faster).
    #    - Skip WC during search if GV is clearly feasible (<= 0.7 * budget).
    #    - Use tight timeouts (0.2s) during search to fail fast on bad cuts.
    #    - Use relaxed timeouts for final verification to ensure stability.
    
    metadata = q.get_metadata()
    num_qubits = metadata.get("num_qubits", 0)
    
    # Valid range: [2, num_qubits]
    n = max(2, int(num_qubits))
    
    # Heuristic: Very small circuits rarely benefit from cutting overhead.
    if n <= 4:
        return n, "GV"

    # Cache: size -> (gv_cost, wc_cost)
    cache = {}

    def get_costs(s, strict=False):
        if s in cache:
            # If we need strict but cached result has 'inf' for WC, we might ideally re-run.
            # But for speed, we rely on the cache unless manually cleared/handled.
            return cache[s]
        
        # Timeouts: Tight for search, longer for final check
        t_out = 0.5 if strict else 0.2
        
        # 1. Check GV
        g_res = compute_gv_cost(q, s, timeout_sec=t_out)
        if isinstance(g_res, tuple):
            g_val = float(g_res[0])
            g_to = bool(g_res[1])
        else:
            g_val = float(g_res)
            g_to = False
        if g_to: g_val = float('inf')

        # 2. Check WC
        # Optimization: Skip WC in search mode if GV is clearly feasible.
        w_val = float('inf')
        check_wc = True
        
        if not strict and g_val <= (float(budget) * 0.7):
            check_wc = False
            
        if check_wc:
            w_res = compute_wc_cost(q, s, timeout_sec=t_out)
            if isinstance(w_res, tuple):
                w_val = float(w_res[0])
                w_to = bool(w_res[1])
            else:
                w_val = float(w_res)
                w_to = False
            if w_to: w_val = float('inf')

        cache[s] = (g_val, w_val)
        return g_val, w_val

    # Binary search for Minimum Feasible Size
    low = 2
    high = n
    best_feasible_size = n 
    
    # Pre-check max size to ensure we have a valid fallback.
    # Max size is usually feasible (cost ~0 or 1).
    g_n, w_n = get_costs(n, strict=False)
    if min(g_n, w_n) > budget:
        return n, ("WC" if w_n < g_n else "GV")

    while low <= high:
        mid = (low + high) // 2
        g, w = get_costs(mid, strict=False)
        
        if min(g, w) <= budget:
            # Feasible. Try smaller.
            best_feasible_size = mid
            high = mid - 1
        else:
            # Infeasible. Need larger.
            low = mid + 1

    final_size = best_feasible_size
    
    # Final Decision Phase
    # Retrieve costs. If we skipped WC or timed out during search, check properly now.
    g_final, w_final = cache.get(final_size, (float('inf'), float('inf')))
    
    # Heuristic: When to re-check WC?
    # If we skipped it (inf) AND GV is expensive (> 50% budget).
    # If GV is cheap, WC is unlikely to be significantly better to justify the cost.
    recheck_wc = False
    if w_final == float('inf') and g_final > (float(budget) * 0.5):
        recheck_wc = True
        
    if recheck_wc:
        w_res = compute_wc_cost(q, final_size, timeout_sec=0.5)
        if isinstance(w_res, tuple):
            val = float(w_res[0])
            if not w_res[1]: w_final = val
        else:
            w_final = float(w_res)
            
    # Select best method based on cost
    final_method = "GV"
    if w_final < g_final and w_final <= budget:
        final_method = "WC"
    elif g_final <= budget:
        final_method = "GV"
    else:
        # Fallback if both over budget
        final_method = "WC" if w_final < g_final else "GV"

    # OE_END
    return final_size, final_method
\end{lstlisting}


\begin{lstlisting}[
  basicstyle=\footnotesize\ttfamily,
  breaklines=true,
  breakatwhitespace=false,
  columns=fullflexible,
  frame=single,
  caption={OpenEvolve Configuration for Multiprogramming},
  label={lst:oe_config_mp},
  captionpos=t,
]
max_iterations: 100
checkpoint_interval: 100
log_level: "INFO"
log_dir: null
random_seed: 42
max_tasks_per_child: 1

diff_based_evolution: false
max_code_length: 30000

early_stopping_patience: null
convergence_threshold: 0.001
early_stopping_metric: score

llm:
  api_base: "${OPENAI_API_BASE}$"
  api_key: "${OPENAI_API_KEY}$"
  primary_model: "${OPENAI_MODEL}$"
  temperature: 0.7
  top_p: 0.9
  max_tokens: 16384
  timeout: 450
  retries: 0
  retry_delay: 5
  reasoning_effort: low

prompt:
  system_message: 
    You are optimizing ONLY the function get_matching_score(self, q1, q2, backend, weighted=False, weights=[]) in evaluation/openevolve_pairing/target.py.
    OUTPUT FORMAT (CRITICAL):
    * Return the full file contents of evaluation/openevolve_pairing/target.py.
    * Output code only: no explanations, no markdown fences, no extra text.
    FUNCTION CONTRACT:
    * Do NOT change the function signature. It MUST remain: get_matching_score(self, q1, q2, backend, weighted=False, weights=[])
    * It MUST return a single float score (higher is better).
    * Keep it deterministic (no randomness or external state).
    METADATA KEYS (from q1.get_metadata()/q2.get_metadata()):
    * depth, num_qubits, num_clbits, num_nonlocal_gates, num_connected_components, number_instructions, num_measurements, num_cnot_gates, program_communication, liveness, parallelism, measurement, entanglement_ratio, critical_depth.
    ARTIFACTS:
    You may receive evaluator artifacts from previous iterations. Use them to infer what kinds of pairs your current scoring function is selecting and how those selections align with strong Pareto rank.
    - rank_distribution_csv: Shows how often each Pareto rank appears overall and within your selected Top-K pairs. Use this to see whether your score is concentrating mass on strong ranks.
    - top_pairs_metrics_csv: Shows the Top-K pairs chosen by your score together with effective utilization, Pareto metric values, Pareto rank, and score. Use this to understand which pair characteristics your current formula is rewarding.
    - top_rank_pairs_all_columns_csv: Expanded pair-level table for top-ranked pairs. Use this to inspect which metadata patterns tend to appear in strong candidates.
    
    CONTEXT:
    The goal is to evolve a scoring function that maximizes the evaluator's objective (avg_rank, corr, or combined) by selecting optimal circuit pairs for multiprogramming based on effective_utilization and fidelity.
  num_top_programs: 1
  num_diverse_programs: 1
  use_template_stochasticity: true
  include_artifacts: true

database:
  db_path: null
  population_size: 100
  archive_size: 20
  num_islands: 3
  migration_interval: 30
  migration_rate: 0.12
  exploration_ratio: 0.65
  exploitation_ratio: 0.3

evaluator:
  timeout: 4800
  max_retries: 1
  parallel_evaluations: 1
  use_llm_feedback: false
\end{lstlisting}

\begin{lstlisting}[
  language=Python,
  basicstyle=\footnotesize\ttfamily,
  numbers=left,
  breaklines=true,
  breakatwhitespace=false,
  columns=fullflexible,
  frame=single,
  caption={Initial Program for Multiprogramming.},
  label={lst:initial_program_mp},
  captionpos=t,
]

def get_matching_score(self, q1, q2, backend, weighted: bool = False, weights = []) -> float:
    """Replacement for Multiprogrammer.get_matching_score."""
    meta1 = q1.get_metadata()
    meta2 = q2.get_metadata()

    depth1 = meta1.get("depth", 0)
    depth2 = meta2.get("depth", 0)
    qubits1 = meta1.get("num_qubits", 0)
    qubits2 = meta2.get("num_qubits", 0)
    nonlocal1 = meta1.get("num_nonlocal_gates", 0)
    nonlocal2 = meta2.get("num_nonlocal_gates", 0)
    cnot1 = meta1.get("num_cnot_gates", 0)
    cnot2 = meta2.get("num_cnot_gates", 0)
    meas1 = meta1.get("num_measurements", 0)
    meas2 = meta2.get("num_measurements", 0)
    instr1 = meta1.get("number_instructions", 0)
    instr2 = meta2.get("number_instructions", 0)
    cc1 = meta1.get("num_connected_components", 0)
    cc2 = meta2.get("num_connected_components", 0)
    liveness1 = meta1.get("liveness", 0.0)
    liveness2 = meta2.get("liveness", 0.0)
    prog_comm1 = meta1.get("program_communication", 0.0)
    prog_comm2 = meta2.get("program_communication", 0.0)
    parallel1 = meta1.get("parallelism", 0.0)
    parallel2 = meta2.get("parallelism", 0.0)
    meas_ratio1 = meta1.get("measurement", 0.0)
    meas_ratio2 = meta2.get("measurement", 0.0)
    ent_ratio1 = meta1.get("entanglement_ratio", 0.0)
    ent_ratio2 = meta2.get("entanglement_ratio", 0.0)
    crit1 = meta1.get("critical_depth", 0.0)
    crit2 = meta2.get("critical_depth", 0.0)

    depth_ratio = min(depth1, depth2) / max(depth1, depth2, 1)
    qubit_ratio = min(qubits1, qubits2) / max(qubits1, qubits2, 1)
    nonlocal_ratio = min(nonlocal1, nonlocal2) / max(nonlocal1, nonlocal2, 1)
    cnot_ratio = min(cnot1, cnot2) / max(cnot1, cnot2, 1)
    meas_ratio = min(meas1, meas2) / max(meas1, meas2, 1)
    instr_ratio = min(instr1, instr2) / max(instr1, instr2, 1)
    instr_diff = abs(instr1 - instr2)
    instr_sum = instr1 + instr2
    instr_diff_norm = instr_diff / max(instr_sum, 1)
    cc_ratio = min(cc1, cc2) / max(cc1, cc2, 1)
    crit_ratio = min(crit1, crit2) / max(crit1, crit2, 1e-9)
    liveness_avg = (liveness1 + liveness2) / 2.0
    prog_comm_avg = (prog_comm1 + prog_comm2) / 2.0
    parallel_avg = (parallel1 + parallel2) / 2.0
    meas_ratio_avg = (meas_ratio1 + meas_ratio2) / 2.0
    ent_ratio_avg = (ent_ratio1 + ent_ratio2) / 2.0
    depth_sim = self.depthComparison(q1, q2)

    util_eff = self.effective_utilization(q1, q2, backend)
    util_eff_norm = util_eff / 100.0
    entanglementDiff = self.entanglementComparison(q1, q2)
    measurementDiff = self.measurementComparison(q1, q2)
    parallelismDiff = self.parallelismComparison(q1, q2)

    # OE_BEGIN
    if weighted and sum(weights) > 0:
        return (
            weights[0] * util_eff +
            weights[1] * entanglementDiff +
            weights[2] * measurementDiff +
            weights[3] * parallelismDiff
        )
    # OE_END
    
    base_score = (util_eff_norm + entanglementDiff + measurementDiff + parallelismDiff) / 4.0

    return base_score
\end{lstlisting}

\begin{lstlisting}[
  language=Python,
  basicstyle=\footnotesize\ttfamily,
  numbers=left,
  breaklines=true,
  breakatwhitespace=false,
  columns=fullflexible,
  frame=single,
  caption={Best Program by Gemini-3-Pro for Multiprogramming},
  label={lst:best_program_mp},
  captionpos=t,
]
def get_matching_score(self, q1, q2, backend, weighted: bool = False, weights = []) -> float:
    """Replacement for Multiprogrammer.get_matching_score."""
    d1 = q1.get_metadata().get("depth", 0)
    d2 = q2.get_metadata().get("depth", 0)
    util = self.effective_utilization(q1, q2, backend) / 100.0
    depth_ratio = min(d1, d2) / max(d1, d2, 1)
    return util * 0.65 + depth_ratio * 0.35
\end{lstlisting}

\begin{figure}[t]
    \centering
    \includegraphics[width=1\linewidth]{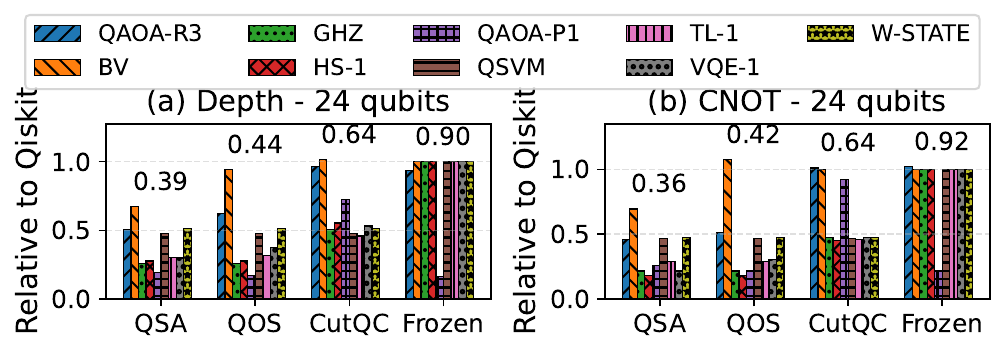}
    \caption{Static metrics comparison for error mitigation on 24-qubit circuits. Lower is better.}
    \label{fig:panel_depth_cnot}
\end{figure}
\section{\name's Impact on Circuit Properties}
\label{appendix:circuit_property}
Figure~\ref{fig:panel_depth_cnot} compares the relative depth and CNOT values of \name and the baselines across circuits of different scales.
Compared to the SOTA QOS, \name consistently reduces both relative depth and relative CNOT by up to 0.06.
These reductions help explain why \name is able to maintain, and in some cases even improve, fidelity.

\section{Additional Ablation Studies}
\label{sec:additional_ablation}

\begin{figure}[h]
    \centering
    \begin{minipage}[t]{0.46\linewidth}
        \centering
        \includegraphics[width=\linewidth]{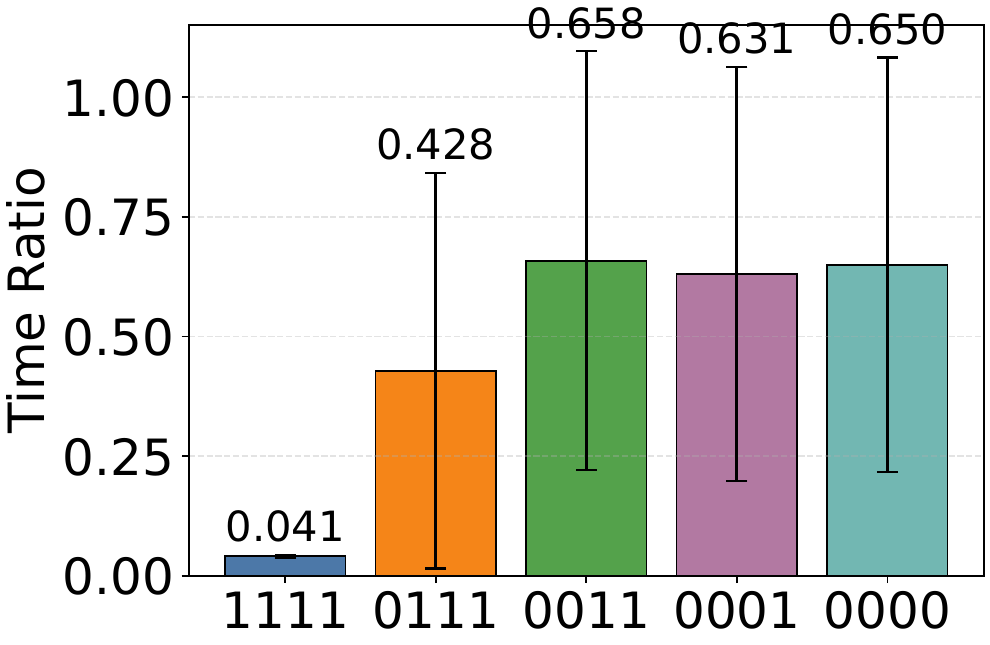}
        \caption{Impact of artifact composition (similar depth/cnot).}
        \label{fig:gem3flash_ablation}
    \end{minipage}
    \hfill
    \begin{minipage}[t]{0.46\linewidth}
        \centering
        \includegraphics[width=\linewidth]{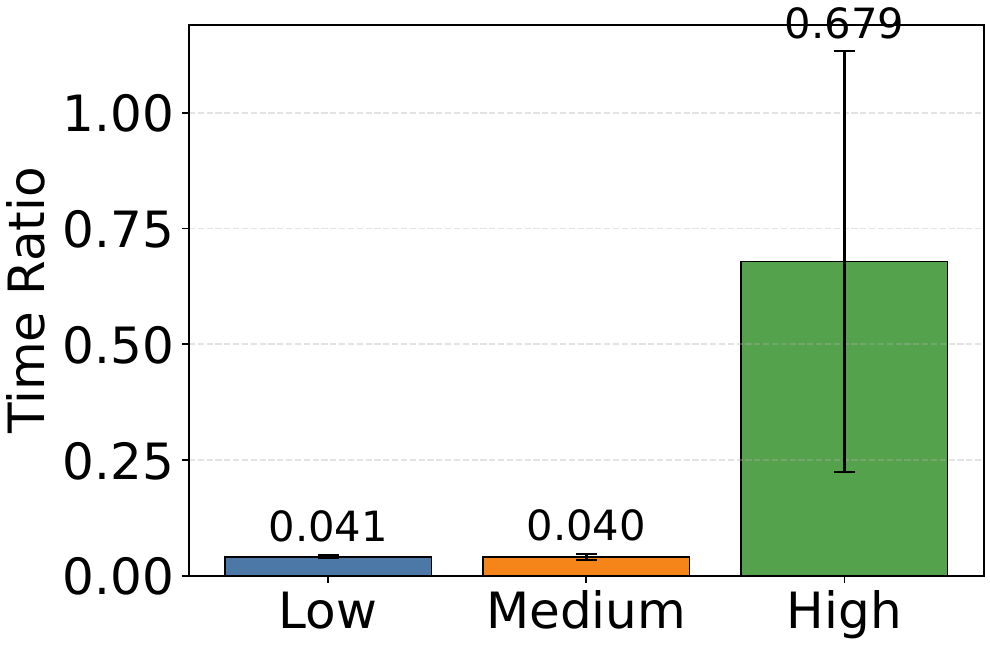}
        \caption{Impact of thinking levels.}
        \label{fig:gem3flash_thinking}
    \end{minipage}
\end{figure}
\noindent\textbf{Ablation of Prompt Composition for Error Mitigation Evolution.}
In Figure~\ref{fig:gem3flash_ablation}, we compare different prompt compositions on Gemini-3-Flash-Preview, the most cost-effective one.
We use four digits to represent the availability of 1) seed program, 2) snapshot (per-circuit), 3) snapshot (summary), 4) static analysis.
When part of the prompt disappears, the mitigation time increases.
The mitigation time plateaus when the seed program and the snapshot (per-circuit) are removed.
This analysis showcases the significance of the prompt components.

\begin{figure}[h]
    \centering
    \begin{minipage}[t]{0.49\linewidth}
        \centering
        \includegraphics[width=\linewidth]{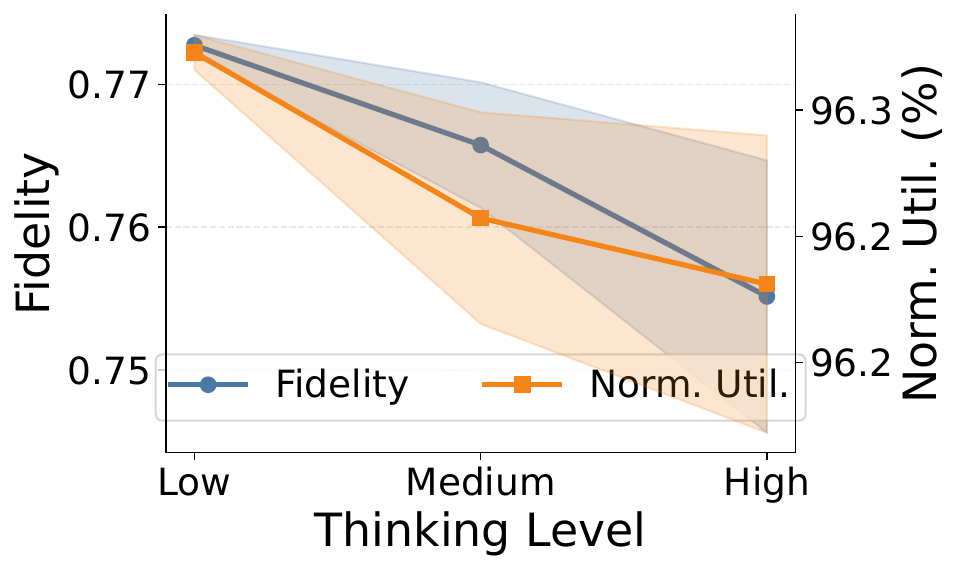}
    \caption{Analysis of Impact of Thinking Level.}
    \label{fig:art_ablation_thinking_level}
    \end{minipage}
    \hfill
    \begin{minipage}[t]{0.49\linewidth}
        \centering
        \includegraphics[width=\linewidth]{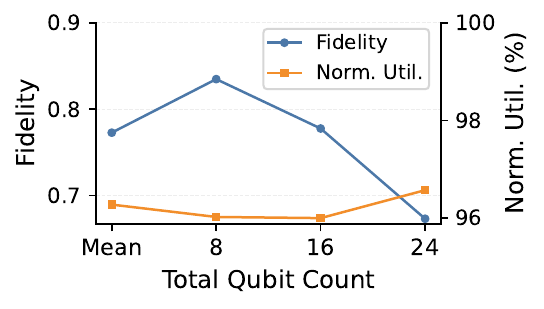}
    \caption{Analysis of Impact of Utils Used in Evolution.}
    \label{fig:art_ablation_util}
    \end{minipage}
\end{figure}

\noindent\textbf{Impact of Thinking Level on Error Mitigation Evolution.}
We analyze the impact of thinking levels on evolution using \GeminiTF by configuring the thinking level in the API.
It turns out that a higher thinking level does not lead to a better result.
Figure~\ref{fig:art_ablation_thinking_level} shows that the fidelity and normalized utilization decreases with the thinking level.
Figure~\ref{fig:gem3flash_thinking} shows that medium thinking level achieves comparable time ratio as the low thinking level while the high thinking level drastically increases the time ratio (lower the better).
We speculate the reason is that the model produces over-complicated solutions at higher thinking levels that fail to make improvements.
In contrast, a low thinking level makes incremental but steady progress in each iteration, which benefits more from the evolution framework.

\noindent\textbf{Impact of Utilizations Used in Multiprogramming Evolution.}
In Figure~\ref{fig:art_ablation_util}, we compare the qubit counts (8,16,24) we use for evolution.
It will impact both the calculation of the target optimization score, and the artifacts.
While single-utilization targets (e.g., 30\%) prioritize specific metrics, adopting the \textit{Mean} across all levels yields the most robust generalization and balanced trade-off between fidelity and utilization.

\begin{figure}[htbp]
  \centering
  \begin{subfigure}[b]{0.49\columnwidth}
    \centering
    \includegraphics[width=\linewidth]{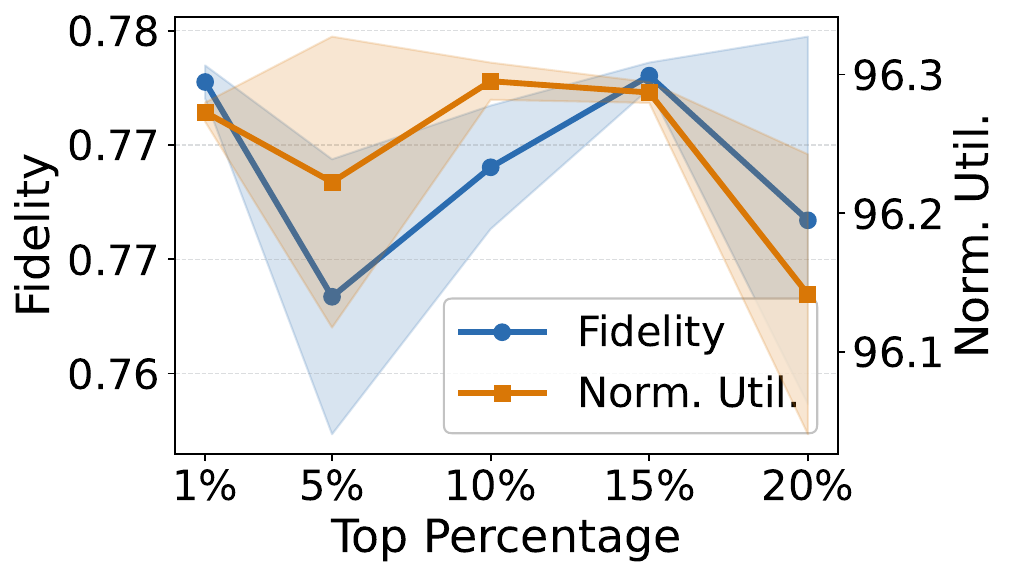}
    \caption{Ablation Studies across the percentage used for evolution.}
    \label{fig:ablation_percentage}
  \end{subfigure}
  \hfill 
  \begin{subfigure}[b]{0.49\columnwidth}
    \centering
    \includegraphics[width=\linewidth]{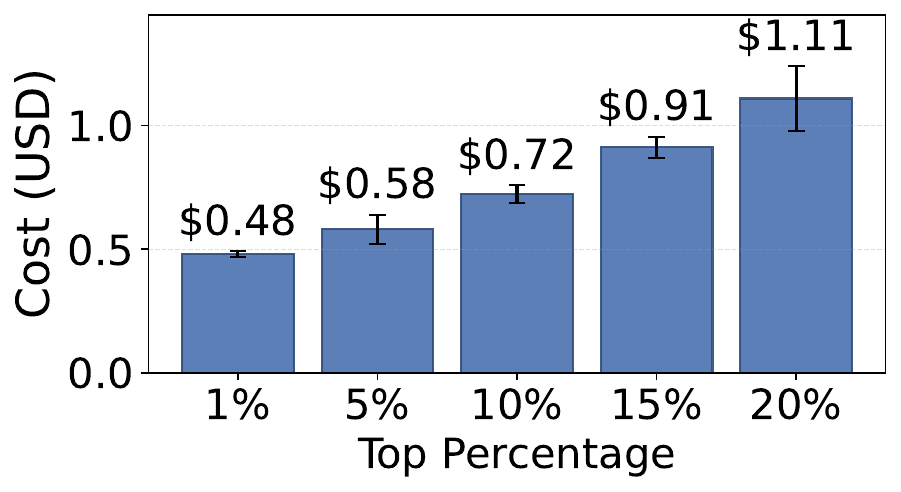}
    \caption{The cost of using different percentage for evolution.}
    \label{fig:ablation_percentage_cost}
  \end{subfigure}
  \caption{Analysis of impact on the percentage used for the evolution.}
  \label{fig:ablation_percentage_combined}
\end{figure}

\noindent\textbf{Impact of Percentage Used in Multiprogramming Evolution.}
We compare the percentage used to compose both the target optimization score, and artifacts during the code evolution. 
As Figure~\ref{fig:ablation_percentage_combined} shows, top 1\% and 5\% percentage can achieve the best fidelity and normalized utilization, while the cost of 1\% is only half the price of 5\% using \GeminiTF. 
In conclusion, from the monetary efficiency point of view, we would pick up the option of 1\% percentage used in evolution.

\section{Fidelity Approximation Details}
\label{appendix:approximation_detail}
Direct fidelity estimation is often computationally infeasible for complex circuits. 
Consequently, these approximation metrics serve as objective functions to surrogate for fidelity.
Table~\ref{tab:proxies_summary} summarizes these fidelity approximation along with their corresponding physical rationales.
We evaluate them as the optimization target in OpenEvolve on Figure~\ref{fig:impact_of_proxy} in the main paper.

\begin{table*}[t]
    \centering
    \small
    \begin{tabularx}{\textwidth}{l X} 
        \toprule
        \textbf{Approximation Metric} & \textbf{Rationale} \\
        \midrule
        Circuit Depth & Represents critical path length; increased depth correlates with decoherence and gate errors. \\
        Instruction Count & Denotes total gate count, reflecting the overall computational load of the quantum circuit. \\
        CNOT Count & Quantifies error-prone entangling gates, which are primary noise sources in NISQ devices. \\
        Non-local Gates & Dictates the requirement for qubit routing and SWAP~\cite{wille2014optimal} insertions, impacting execution efficiency. \\
        Measurement & Reflects measurement operation count, contributing to readout errors and temporal overhead. \\
        \bottomrule
    \end{tabularx}
    \caption{Summary of Performance Proxies in \name.}
    \label{tab:proxies_summary}
\end{table*}

\section{Optimization Baseline Details}
\label{app:optimization_baselines}

We compare \name against three conventional optimization strategies to
determine whether its gains can be achieved by optimizing the same
error-mitigation decisions without evolving program logic.
All methods choose from the same output decision space: the target circuit fragment size and the cutting method (GV or WC).
They are evaluated on the same workloads using qiskit quantum simulator, under a 60\,s search deadline.

\noindent\textbf{Random.}
Random search uniformly samples valid fragment-size and cutting-method
configurations and retains the best configuration observed within the deadline.

\noindent\textbf{Bayesian Optimization.}
BO treats fragment size and cutting method as optimization variables and
sequentially selects configurations using a constrained Gaussian-process
surrogate.
After evaluating each configuration, the observed reward is used to update the
surrogate and select the next candidate until the deadline is reached.

\noindent\textbf{Reinforcement Learning.}
RL learns a feature-based policy that maps circuit characteristics to a
fragment size and cutting method.
We train the policy using the same circuit coreset used by \name, so that both
methods receive comparable workload coverage during optimization.
At deployment, the learned policy directly selects an action from this fixed
feature--action representation.

\noindent\textbf{Ours.}
Unlike these baselines, \name does not optimize only the final configuration.
Instead, the LLM mutates the underlying error-mitigation program and can change
the control logic used to explore fragment sizes and cutting methods.
For example, the evolved program can alter the search order, reuse intermediate
results, or terminate unproductive searches early, while remaining subject to
the same execution deadline.
This comparison therefore isolates the benefit of program-level evolutionary
optimization from conventional search over predefined decisions.

\section{Additional Discussions}
\label{appendix:additional_discussion}

\noindent\textbf{Robustness and limitations.}
\name's robustness to hardware calibration drift depends on the metrics. Mitigation time and normalized utilization are mainly determined by the evolved software policy running on classical computers and are therefore less sensitive to calibration changes. In contrast, fidelity may vary with the hardware noise level, and higher the noise, lower the fidelity gain. Regarding new workloads, our experimental results show that \name can generalize to existing topologies with varying scales. 

\noindent\textbf{Extension to quantum error correction.}
For quantum error correction, \name can evolve decoder implementations or decoding schedules. Weakness analysis can identify the code paths that dominate logical error rate or decoding latency, while the reward can directly combine these metrics with resource overhead. The evolution loop can then search for policies that improve decoding quality without violating real-time latency constraints.

\end{document}